\documentclass[reprint,superscriptaddress,aps,prx,nofootinbib]{revtex4-2}
\usepackage[T1]{fontenc}
\usepackage[utf8]{inputenc}
\usepackage[normalem]{ulem}
\usepackage[cal=boondoxo]{mathalfa}
\usepackage[breaklinks,colorlinks,urlcolor=blue,linkcolor=blue,anchorcolor=blue,citecolor=blue]{hyperref}
\usepackage{lmodern,amsmath,amssymb,graphicx,xcolor,comment,braket,enumerate,verbatim,float,dsfont,bm}

\usepackage{silence}
\makeatletter
\newcommand*{\balancecolsandclearpage}{%
  \close@column@grid
  \clearpage
  \twocolumngrid
}
\makeatother

\renewcommand\footnoterule{}
\expandafter\let\csname equation*\endcsname\relax
\expandafter\let\csname endequation*\endcsname\relax
\def\ii{{\rm i}}  
\def\dpp{\boldsymbol{\wp}}

\def\rb{{\bf r}}

\def\sp{\hat{\sigma}^+}  
\def\sm{\hat{\sigma}^-} 
\def\sz{\hat{\sigma}^z}

\def\bra#1{\mathinner{\langle{#1}|}}
\def\ket#1{\mathinner{|{#1}\rangle}}
\def\braket#1{\mathinner{\langle{#1}\rangle}}

\newcommand{\rr}{\mathbf{r}}

\newcommand{\avg}[1]{\ensuremath{\langle #1 \rangle}}
\newcommand{\pare}[1]{\left( {#1} \right)}

\newcommand{\cpare}[1]{\left\{ {#1} \right\}}

\newcommand{\new}[1]{\textcolor{black}{#1}}
\newcommand{\Tr}{\text{Tr}}

\newcommand\norm[1]{\left\lVert#1\right\rVert}
\newcommand{\w}{\omega}

\newcommand{\kk}{\mathbf{k}}

\newcommand{\nn}{\boldsymbol{n}}
\newcommand{\spl}{\hat \sigma^{+}}
\newcommand{\smi}{\hat \sigma^{-}}

\newcommand{\qq}{\mathbf{q}}

\renewcommand\footnotemark{}

\usepackage{natbib}
\usepackage[resetlabels]{multibib}
\newcites{supp}{Supplementary References}
\begin{document}
%\newrefsection
\title{Universal scaling laws for correlated decay of many-body quantum systems}
\author{Wai-Keong Mok}
\thanks{These authors contributed equally to this work.}
\affiliation{Institute for Quantum Information and Matter, California Institute of Technology, Pasadena, CA 91125, USA}
\author{Avishi Poddar}
\thanks{These authors contributed equally to this work.}
\affiliation{Department of Physics, Columbia University, New York, NY 10027, USA}
\author{Eric Sierra}
\affiliation{Department of Physics, Columbia University, New York, NY 10027, USA}
\author{Cosimo C. Rusconi}
\affiliation{Department of Physics, Columbia University, New York, NY 10027, USA}
\author{John Preskill}
\affiliation{Institute for Quantum Information and Matter, California Institute of Technology, Pasadena, CA 91125, USA}
\affiliation{AWS Center for Quantum Computing, Pasadena CA 91125}
\author{Ana Asenjo-Garcia}
\email{ana.asenjo@columbia.edu}
\affiliation{Department of Physics, Columbia University, New York, NY 10027, USA}

\date{\today}
\begin{abstract}
Quantum systems are open, continually exchanging energy and information with the surrounding environment. This interaction leads to decoherence and decay of quantum states. In complex systems, formed by many particles, decay can become correlated and enhanced. A fundamental question then arises: what is the maximal decay rate of a large quantum system, and how does it scale with its size? In this work, we address these issues by reformulating the problem into finding the ground state energy of a generic spin Hamiltonian. Inspired by recent work in Hamiltonian complexity theory, we establish rigorous and general upper and lower bounds on the maximal decay rate. These bounds are universal, as they hold for a broad class of Markovian many-body quantum systems. For many physically-relevant systems, the bounds are asymptotically tight, resulting in exact scaling laws with system size. Specifically, for large atomic arrays in free space, these scalings depend only  on the arrays' dimensionality and are insensitive to details at short length-scales.
The scaling laws set fundamental limits on the decay rates of all quantum states, \new{shed light on the behavior of generic driven-dissipative systems, and may ultimately constrain the scalability of quantum processors and simulators based on atom arrays.}
\end{abstract}
\maketitle

Understanding the quantum dynamics of far-from-equilibrium open many-body systems is a major frontier in physics.  From a fundamental perspective, the interplay between energy pumping and dissipation allows for the emergence of phases that transcend the paradigms established by equilibrium statistical physics. Examples in quantum optics include the superradiant laser~\cite{Haake1993superradiant,Meiser2009,bohnet} and the driven Dicke phase transition~~\cite{narducci,carmichael_1980,Ferioli2023nonequilibrium}. From an applied standpoint, the full potential of quantum technologies -- including quantum computing, quantum simulation, and metrology -- is realized only with large systems that remain coherent despite their coupling to a bath.

In systems formed by many particles, the always-present vacuum fluctuations mediate long-range dissipative interactions that cannot be switched off, inducing correlated decay that may increase with system size. Such decay processes are collectively enhanced if the particles are tightly packed. Correlated decay may thus become the ultimate source of decoherence for many quantum technologies. For instance, it may alter the signal-to-noise ratio in metrology experiments such as atomic clocks or spin squeezing. Similarly, in large-scale quantum computers, it can lead to much shorter coherence times than the predicted timescales using independent noise models and may hinder quantum error correction~\cite{Nielsen2010quantum,Preskill2018quantum, lemberger2017effect}. On the other hand, correlated decay is a critical requirement for other applications, such as the development of new light sources~\cite{Meiser2009,bohnet,Gonzalez2015deterministic}, the dissipative preparation of correlated many-body states~\cite{Orioli2022emergent}, or the protection of logical quantum information via dissipation~\cite{Leghtas2015confining, Gertler2021protecting}.

Due to the exponential complexity associated with large quantum systems, exactly computing the largest decay rate is a formidable challenge. This problem remains unsolved except in trivial cases, such as permutationally-symmetric models (e.g., atoms coupled to a cavity) and non-interacting systems.  In generic situations, finding the largest decay rate is as difficult as determining the ground state of a general $2$-local Hamiltonian, which is known to be a QMA-complete problem -- expected to be hard even for a quantum computer~\cite{Kempe2006complexity}. This complexity is compounded by the diversity of experimental platforms, with many candidates serving as qubits (neutral atoms, molecules, ions, superconducting qubits, quantum dots, vacancy centers, among others) as well as propagators of the interactions between them (electromagnetic field, and other bosonic collective excitations such as phonons, magnons, etc).

In this work, we find upper and lower bounds to the maximal decay rate by leveraging tools from Hamiltonian complexity theory~\cite{Gharibian2015quantum,Osborne2012hamiltonian, bravyi2019approximation} and applying them in the context of out-of-equilibrium quantum dynamics. For a large class of systems, these bounds are asymptotically tight, thus yielding scaling laws with system size that only depend on the spectral properties of the  decoherence matrix $\mathbf{\Gamma}$, whose dimension is linear in system size. The bounds are obtained by means of product-state ansatzes, and thus imply that entanglement does not play any role in the scaling. \new{Our results are formally rigorous and do not rely on mean field approximations: while the bounds are derived using product states, they remain valid for any state (including entangled states), thus going beyond previous approaches based on mean field theory~\cite{Gross1982} or approximate numerical methods~\cite{rubies-bigorda2023cumulant,Mink2023,Masson2022NatComm}.} We apply these tools to the specific case of ordered atomic arrays~\cite{Endres2016,Barredo2016, Kim2016,Kumar2018,kaufman} and lattices in free space~\cite{bloch}, which have become an all-around platform for different quantum technologies, ranging from quantum computing~\cite{Bluvstein2024} and quantum simulation~\cite{Bakr2010, bernien,scholl} to atomic clocks~\cite{madjarov,norcia,Hutson2024} and spin squeezing~\cite{Bornet2023, eckner, franke}. In the physically-relevant regime of lattice constants similar to the resonance wavelength, the maximal decay rate scales as $\sim N^{\frac{3}{2}-\frac{1}{\text{2D}}}$, where D is the array dimensionality. This scaling law is universal as it does not depend on specific details of the array such as lattice geometry or atomic polarization and has implications in a broad set of problems ranging from quantum dynamics to metrology and quantum computation.

\begin{figure}
\includegraphics[width=\columnwidth,clip,keepaspectratio]{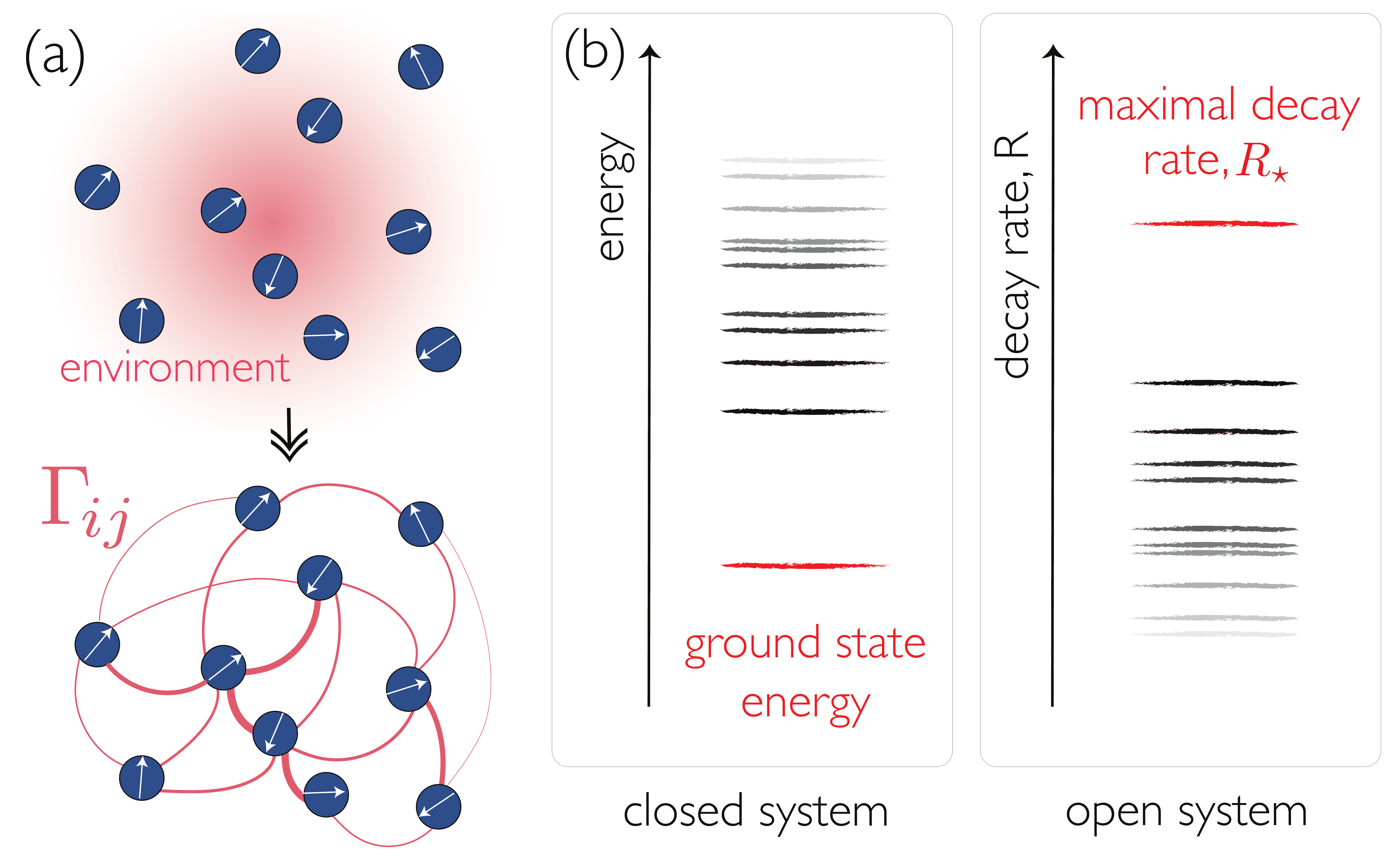}
\caption{Generic qubit ensemble described as an out-of-equilibrium, open, many-body quantum system. (a) In Markovian baths, integrating out the environment degrees of freedom yields a spin model with coherent and dissipative interactions. The dissipative couplings between $N$ qubits are given by the decoherence matrix $\mathbf{\Gamma} = (\Gamma_{ij})_{i,j=1}^{N}$. (b) For a closed system, the ground state is the state of minimal energy. For an open system, finding the state with maximal decay rate ($R_\star$) is analogous to finding the ground state energy of a Hamiltonian.}\label{fig1}
\end{figure}

\textbf{Theory background.} A broad class of Markovian many-body open quantum systems of $N$ qubits [Fig.~\ref{fig1}(a)] is described by the Lindblad master equation
\begin{equation}
    \dot{\hat{\rho}} = -\frac{i}{\hbar}[\hat{H},\hat{\rho}] + \sum_{i,j=1}^{N} \Gamma_{ij} \left(\sm_i \hat{\rho} \sp_j - \frac{1}{2}\{\sp_j \sm_i, \hat{\rho} \} \right),
\label{eq:lindbladME}
\end{equation}
where $\hat{\sigma}_i^\pm = (\hat{\sigma}_i^x \pm i\hat{\sigma}_i^y)/2$ are the raising and lowering operators for qubit $i$. In the above equation, $\hat{H}$ is an arbitrary qubit Hamiltonian that commutes with the total excitation operator $\hat{n}_\text{exc}=\sum_i \hat{\sigma}^+_i\hat{\sigma}^-_i$, while the dissipative interactions are represented by the \new{Hermitian} decoherence matrix $\mathbf{\Gamma} = (\Gamma_{ij})_{i,j=1}^{N}$. \new{The master equation can be generalized significantly beyond Eq.~\eqref{eq:lindbladME} to include arbitrary local and multi-qubit Hamiltonian interactions, coherent driving, decoherence terms (such as incoherent driving, coupling to a finite-temperature reservoir, dephasing), as well as disorder in $\mathbf{\Gamma}$, without affecting the conclusions of this paper [see Sec.~\ref{sec:disorderandmore} of the Supplementary Information (SI) for a detailed discussion].}

For the master equation to describe a physically valid evolution (i.e., a completely positive and trace preserving map), $\mathbf{\Gamma}$ must be positive semidefinite (i.e., $\mathbf{\Gamma} \succeq 0$)~\cite{Nielsen2010quantum}. This ensures non-negative eigenvalues $\{\Gamma_\mu\}$, which are physically interpreted as collective transition rates. Since $\mathbf{\Gamma} \succeq 0$, the spectral norm $\norm{\mathbf{\Gamma}}$ is equal to $\Gamma_{\text{max}}$, the largest collective transition rate (i.e., the largest eigenvalue of the matrix). \new{We define $\Gamma_0 = \sum_{i=1}^{N}\Gamma_{ii}/N$ to be the average individual decay rate. }

The instantaneous decay rate of the many-body system, $R$, is computed as the expectation value of an ``auxiliary'' (and Hermitian) Hamiltonian $\hat{H}_\Gamma$~\cite{mok2023dicke}, i.e., 
\begin{equation}\label{eq:Rate}
    R = -\frac{d}{dt} \braket{\hat{n}_{\text{exc}}}\equiv \braket{\hat{H}_\Gamma}/\hbar 
\end{equation}
where 
\begin{equation}
    \hat{H}_\Gamma =\hbar \sum_{i,j=1}^{N} \new{\Gamma_{ji}} \sp_i \sm_j = \hbar\sum_{\mu = 1}^N \Gamma_\mu \hat{c}_\mu^\dag \hat{c}_\mu,
\label{eq:haux}
\end{equation}
\new{where $\Gamma_{ji}=\Gamma_{ij}^*$.} The last equality is achieved by means of collective jump operators $\hat{c}_\mu = \sum_{i=1}^{N} \alpha_i^{(\mu)} \sm_i$, with \new{$\vec{\alpha}^{(\mu)}$} being the \new{normalized} eigenvectors of $\mathbf{\Gamma}$ (\new{the largest eigenvalue is} $\Gamma_{\text{max}}\equiv\Gamma_1$). Generically, the ``auxiliary'' Hamiltonian $\hat{H}_\Gamma$ describes an $\text{XY}$ model defined on a weighted interaction graph with a local transverse field.  In the specific case where the interactions are mediated by the electromagnetic field, $\mathbf{\Gamma}$ is proportional to the vacuum's Green's function~\cite{Lehmberg1970a} and the decay rate is exactly equal to the integrated photon emission rate over all emission angles. 

\textbf{Lower and upper bounds.} Our goal is to set theoretical limits on the maximal decay rate $R_\star$, \new{the maximum value of $R$ over all possible many-body quantum states. As we demonstrate in Sec.~\ref{app:Bound_Observable} of the SI, $R_\star$ also sets bounds on the rates of change for general observables. For instance, we can bound $|d\braket{\hat{A}}/dt| \leq R_\star \lVert\hat{A}\rVert$ for an arbitrary positive operator $\hat{A}$, assuming $\lVert\hat{H}\rVert \ll R_\star$. Complementary bounds for generic local observables are discussed in the SI.} Computing $R_\star$ amounts to calculating the spectral radius of $\hat{H}_\Gamma$ (since $\hat{H}_\Gamma \succeq 0$), or equivalently the ground state energy of $-\hat{H}_\Gamma$, as depicted in Fig.~\ref{fig1}(b). Finding the exact energy is expected to be hard~\cite{Kempe2006complexity}, except in two limiting cases. For non-interacting qubits (with $\Gamma_{ij} =\Gamma_0\delta_{i,j}$), $R_\star = N\Gamma_0$. In the Dicke limit (i.e., with all-to-all interactions such that $\Gamma_{ij} = \Gamma_0 \;\forall \,i, j$), $R_\star = N(N+2)\Gamma_0/4$~\cite{dicke1954coherence,Gross1982}. These two cases serve as trivial lower and upper bounds, respectively, for $R_\star$ in arbitrary environments. Below, we derive tighter bounds.

\textit{Lower bound from a variational ansatz.} The canonical way to obtain lower bounds on $R_\star$ (or equivalently, upper bounds on the ground state energy of $-\hat{H}_\Gamma$) is to use a variational ansatz for a trial wavefunction. We choose the product state ansatz $\ket{\psi} = (\cos(\theta/2) \ket{g} + \sin(\theta/2)\ket{e})^{\otimes N}$, for which the decay rate is found to be
\begin{equation}
    R_\psi(\theta) = \frac{1}{\hbar}\braket{\psi|\hat{H}_{\Gamma}|\psi}= \frac{N\Gamma_0}{2} (1-\cos \theta) + \frac{S}{4} \sin^2 \theta,
\label{eq:Rpsi_productstate}
\end{equation}
where $S\equiv \sum_{i \neq j} \Gamma_{ij}$ is the sum of dissipative interactions in the system. The maximal value of $R_{\psi} = (N\Gamma_0+S)^2/(4S)$ is attained for the mixing angle $\cos \theta = -N\Gamma_0/S$ if $S \geq N\Gamma_0$, and $\text{max}_\psi R_\psi = N\Gamma_0$ otherwise. Given that $-N\Gamma_0 \leq S \leq N(N-1)\Gamma_0$ is necessary to satisfy $\mathbf{\Gamma} \succeq 0$~\cite{marsli2015bounds}, our lower bound is consistent with the trivial upper bound from the Dicke model. \new{In the simple case where $\Gamma_{ij} \geq 0$, $R_\star \propto S$ (see Sec.~\ref{app:inphase} of the SI). In the following, we do not make such an assumption.}

Modifying the ansatz to include locally-dependent relative phases between $\ket{e}$ and $\ket{g}$ based on the dominant eigenvector of $\mathbf{\Gamma}$ and fixing the excitation density to $1/2$ yields an alternative lower bound (see Sec.~\ref{app:translation} of the SI), 
\begin{equation}
    R_\star = \max_{\ket{\psi}} \sum_\mu \Gamma_\mu \new{\braket{\psi|\hat{c}_\mu^\dag \hat{c}_\mu|\psi}} \geq \Gamma_{\text{max}}\lVert\hat{c}_1^\dag \hat{c}_1\rVert^2\geq \frac{N \Gamma_{\text{max}}}{4(\Delta^2 + 1)},
    \label{eq:product_lower}
\end{equation}
where $0 \leq \Delta \leq \sqrt{N-1}$ is the relative fluctuation of the entries of the dominant eigenvector of $\mathbf{\Gamma}$. This gives a tighter lower bound if the decay is delocalized (i.e., if the brightest collective jump operator has approximately uniform spatial support over all qubits), characterized by the regime where $\Delta = O(1)$. In particular, for a translationally-invariant system, $\Delta = 0$. \new{We remark that bounds showing a similar scaling can be derived using more general ansatzes. In Sec.~\ref{appendix:lb} of the SI, we obtain a lower bound with the same scaling as Eq.~\eqref{eq:product_lower} using entangled states.}

\textit{Upper bound from \new{approximation theory}.} Our first main result is to find an asymptotically tighter upper bound for $R_\star$ by harnessing well-established theoretical guarantees for product state approximations. For many physically relevant systems, this gives us the exact scaling for $R_\star$ with system size. Let us write $\hat{H}_{\Gamma} = \hat{H}_{\text{diag}} + \hat{H}_{\text{XY}}$, where $\hat{H}_{\text{diag}} =\hbar\sum_{i=1}^N \new{\Gamma_{ii}} \hat{\sigma}_i^+ \hat{\sigma}_i^-$ and
\begin{equation}
    \hat{H}_{\text{XY}} = \frac{\hbar}{4} \sum_{i \neq j}  \Gamma_{ij} (\hat{\sigma}_i^x \hat{\sigma}_j^x + \hat{\sigma}_i^y \hat{\sigma}_j^y).
\label{eq:Hxy}
\end{equation}
By the triangle inequality, \new{$R_\star \leq N \Gamma_0 + \Vert\hat{H}_{\text{XY}}\Vert /\hbar$}. Since $\hat{H}_{\text{XY}}$ is $2$-local and traceless, we employ a recent result from Bravyi \textit{et al.}~\cite{bravyi2019approximation} to write 
\begin{equation}
    \Vert\hat{H}_\text{XY}\Vert \leq 6 \hbar \,R_{\text{prod}}(\hat{H}_{\text{XY}}),
\label{eq:bravyi_bound}
\end{equation}
where $R_{\text{prod}}(\hat{H}_{\text{XY}})$ is the best product state approximation to $\Vert \hat{H}_{\text{XY}}\Vert/\hbar$. Restricting to product states, $\hat{H}_{\text{XY}}$ reduces to a classical $\text{XY}$ Hamiltonian
\begin{equation}
    H_{\text{XY}} = \frac{\hbar}{4} \sum_{i\neq j}\Gamma_{ij} \vec{s}_i \cdot \vec{s}_j = \frac{\hbar}{4} \text{Tr}(\mathbf{\tilde{\Gamma}} \mathbf{\Sigma}),
\label{eq:classical_XY}
\end{equation}
with $\vec{s}_i \in \mathbb{R}^2$, $\norm{\vec{s}_i} \leq 1$. In the above equation, $(\mathbf{\Sigma})_{ij} = \vec{s}_i \cdot \vec{s}_j$ is the Gram matrix for the vectors $\{\vec{s}_i\}$ and $\mathbf{\tilde{\Gamma}}$ is the off-diagonal matrix $\mathbf{\Gamma} - \Gamma_0\mathbf{I}_N$. By means of the inequality $\text{Tr}(\tilde{\mathbf{\Gamma}} \mathbf{\Sigma}) \leq \lVert\mathbf{\tilde{\Gamma}}\rVert \text{Tr} (\mathbf{\Sigma})$~\cite{lasserre1995trace}, we obtain
\begin{equation}
    R_{\text{prod}}(\hat{H}_{\text{XY}}) \leq \frac{N}{4} \,(\Gamma_{\text{max}} - \Gamma_0).
\end{equation}

Combining this inequality with Eqs.~\eqref{eq:bravyi_bound} and~\eqref{eq:product_lower}, we find the general bounds
\begin{equation}
   \text{max}\left\{ N\Gamma_0, \frac{N\Gamma_{\text{max}}}{4(\Delta^2 + 1)}\right\} \leq R_\star \leq \frac{N}{2} \left(3 \Gamma_{\text{max}} - \Gamma_0\right).
\label{eq:master_bound}
\end{equation}
For the upper bound, equality is achieved for non-interacting qubits (such that $\Gamma_{\text{max}} = \Gamma_0$). Equation~\eqref{eq:master_bound} also implies that for `sufficiently weak' interactions (such that $\Gamma_{\text{max}}$ is asymptotically independent of $N$), the maximal decay rate scales only linearly with system size. This generalizes some of the authors' recent results on the  impossibility of Dicke superradiance with nearest-neighbor interactions~\cite{mok2023dicke}, to systems with arbitrary interaction range and geometry. 

While the use of product states is not strictly needed to obtain an upper bound on $R_\star$, it provides several advantages. Physically, it implies that entanglement is not necessary for a system to dissipate at a rate near the theoretical maximum scaling, complementary to previous observations about the role of entanglement in spontaneous transient superradiance~\cite{wolfe2014certifying,bojer2022dicke,lohof2023signatures}. \new{We stress that our use of product states here is fundamentally different from many previous analytical studies based on mean-field theory, since the state decaying at rate $R_\star$ is, in general, highly entangled.}

\textbf{Universal scaling laws.} Our bounds are tight for systems with delocalized decay (differing only by a constant factor), thus yielding scaling laws for the maximal decay rate. Taking $\Delta = O(1)$ in Eq.~\eqref{eq:master_bound}, we find
\begin{equation}\label{eq:superlaw}
R_\star\sim N\Gamma_\text{max},
\end{equation}
which is one of the main results of this paper. \new{Despite its apparent simplicity,} the scaling law $R_\star \sim N \Gamma_{\text{max}}$ in the delocalized regime is non-trivial and certainly not true for arbitrary systems. More broadly, in Sec.~\ref{app:reductio} of SI we prove that there are no general scaling laws on $R_\star$ that depend solely on system size and the spectrum of $\mathbf{\Gamma}$. Since $\Gamma_{\text{max}}$ can be computed numerically in $O(N^3)$ time, the scaling law provides an efficient scheme to approximate $R_\star$ for large system sizes with quasi translation invariance [i.e., such that $\Delta = O(1)$]. As discussed in Sec.~\ref{sec:disorderandmore} of SI, these scaling laws are robust to disorder in $\mathbf{\Gamma}$, and hold even in the presence of any number of local Hamiltonian and dissipative terms. \new{Our scaling law} reveals important insights about the $N^2$ scaling in \new{Dicke superradiance}: one factor of $N$ arises from the permutation symmetry, and the other from the delocalized nature of the dominant decay channel together with a non-vanishing excitation density at large $N$. 

It may seem surprising that a product state yields the same asymptotic decay rate as the entangled Dicke state, but this can be thought of as an instance of the quantum de Finetti theorem~\cite{Renner,caves2002unknown}: Since $\hat{H}_\Gamma$ is $2$-local, it suffices to only consider the two-body reduced density matrix of the permutationally symmetric Dicke state, which is close to a product state (with trace distance vanishing as $1/N$). Our results show that the accuracy of the mean field (product state) ansatz holds more generally, even when the permutation symmetry is broken.  

\textbf{Maximal decay rate of atomic arrays in free space.}
We now focus on ordered lattices of two-level atoms in free space, whose interactions are described by the propagator of the electromagnetic field evaluated at the resonance frequency $\omega_0$, which is a long-ranged function with oscillating sign (see Sec.~\ref{app:scaling_Gamma_max} of SI). This makes the problem of finding the ground state of $-\hat{H}_\Gamma$ non-trivial, and is thus a perfect candidate to showcase the strength of our theoretical tools. Nevertheless, our formalism is not restricted to electric-dipole-mediated interactions in free space, but can also describe magnetic-dipole or electric-quadrupole interactions in arbitrarily complex dielectric structures.

In the large $N$ limit, and for a large range of lattice constants, the functional dependence on system size of the largest transition rate is only determined by the dimensionality of the array~\cite{sierra2022dicke}. One can relate the scaling with $N$ to the presence of divergences of $\Gamma(\textbf{k})$ in reciprocal space as $|\textbf{k}|$ approaches $k_0\equiv\omega_0/c$. Divergences do not occur for one-dimensional (1D) arrays. They appear for two- and three-dimensional (2D, 3D) lattices, as the number of atoms per volume increases, enhancing constructive interference of photon emission for certain wavevectors. For a D-dimensional array, the largest transition rate scales as $\Gamma_\text{max}^{\text{(D)}}/\Gamma_0 \sim (k_0 d)^{-\frac{\text{D+1}}{2}} N^{\frac{\text{D-1}}{\text{2D}}}$. \new{This result can be easily generalized to compute the scaling of $\Gamma_\text{max}$ for an arbitrary $\text{D}$-dimensional lattice in $\delta$-dimensional free space, with $\text{D}\leq\delta$ (see Sec.~\ref{app:Higher_Dimensions}). For 3D free space, the associated jump operator is delocalized (see Secs.~\ref{app:scaling_Gamma_max} and \ref{sec:Delocalized_Decay} of SI). Accordingly, the scaling in Eq.~(\ref{eq:superlaw}) holds for the case of atomic arrays in free space.}

The asymptotic scaling of the maximal decay rate depends on the array dimensionality (D $\in \{1,2,3\}$) as
\begin{equation}
\frac{R_\star^{\text{(D)}}}{\Gamma_0} \sim N^{\frac{3}{2}-\frac{1}{2\text{D}}}.
\label{eq:array_scaling_law}
\end{equation}
This expression, which is one of the main results of the paper, is universal in the sense that it does not depend on microscopic details (such as lattice constant, geometry, polarization), which only appear as prefactors that do not change the scaling as long as the atom number is large enough. \new{We investigated the effects of common experimental imperfections such as position disorder and finite temperature on the scaling laws. In Sec.~\ref{app:Disorder} of SI, we prove that the upper bound on $R_\star$ remains robust against these imperfections, and provide numerical evidence indicating that the scaling of the lower bound is preserved.} The scaling in Eq.~\eqref{eq:array_scaling_law} differs significantly from that expected in the Dicke limit as $N \to \infty$. The departure is largest for 1D arrays, whose largest decay rate effectively scales as that of a collection of non-interacting atoms. 

\begin{figure}[!t]
\includegraphics[width=\columnwidth,clip,keepaspectratio]{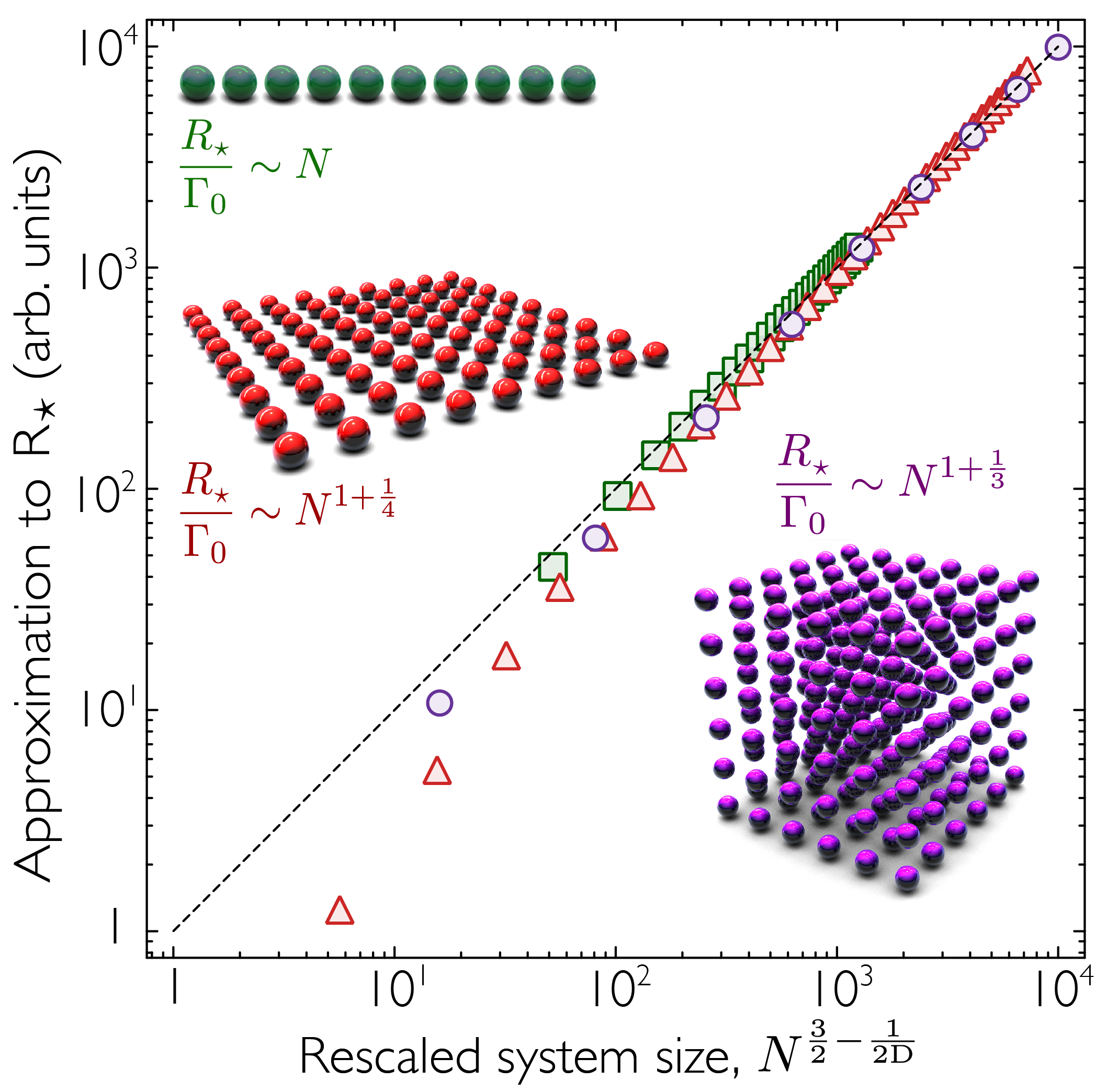}
\caption{Scaling with system size of the numerical approximation for the maximal decay rate $R_\star$ given by the SDP solution, for different lattice dimensionalities with lattice constant $d=0.4\lambda_0$. For 1D ({\scriptsize $\square$}) and 2D ({\scriptsize $\triangle$}) arrays, the atoms are perpendicularly polarized; for the 3D ($\circ$) lattice, the atoms are polarized along one axis of the array. \new{The dashed black line is a guide to the eye representing the analytical scaling law $R_\star\sim N^{\frac{3}{2}-\frac{1}{\text{2D}}}\Gamma_0$}.} \label{fig2}
\end{figure}

We benchmark our analytical scaling laws via a semidefinite program (SDP) relaxation~\cite{Boyd2004convex}, which provides an upper bound to $R_{\text{prod}}(\hat{H}_{\text{XY}})$ [see Eqs.~\eqref{eq:bravyi_bound} and~\eqref{eq:classical_XY}].  SDP relaxations have been used to lower-bound different types of ground state problems~\cite{barthel2012solving,baumgratz2012lower,baccari2020verifying,parekh2021application} and, more recently, ground-state observables~\cite{Wang2024}. The SDP relaxation to Eq.~\eqref{eq:classical_XY} reads
\begin{equation}
\begin{split}
    \max_{\mathbf{X}\succeq 0, \mathbf{X}^T = \mathbf{X}} \quad & \frac{1}{4} \text{Tr}(\mathbf{\tilde{\Gamma}} \mathbf{X}) \\ \text{subject to} \quad & \mathbf{X}_{ii} \leq 1 \quad \forall i = 1,\ldots,N
\end{split}
\label{eq:sdp}
\end{equation}
which can be solved in polynomial time. Here, $\mathbf{X}$ is the Gram matrix with elements $\mathbf{X}_{ij} = \vec{x}_i \cdot \vec{x}_j$, $\vec{x}_i \in \mathbb{R}^N$. This is a relaxation since the Gram matrix $\mathbf{\Sigma}$ associated to $R_{\text{prod}}(\hat{H}_{\text{XY}})$ has a rank of at most $2$, while $\mathbf{X}$ can have a rank of up to $N$. Relaxing the rank constraint renders the optimization problem convex, and thus efficiently solvable. Using an SDP solver~\cite{diamond2016cvxpy}, we obtain a good approximation of $R_\star$ as shown in Fig.~\ref{fig2} for arrays with lattice constant $d=0.4\lambda_0$, where $\lambda_0=2\pi/k_0$ is the wavelength associated to the resonant transition.  

\begin{figure*}[!th]
    \centering
    \includegraphics[width=\textwidth]{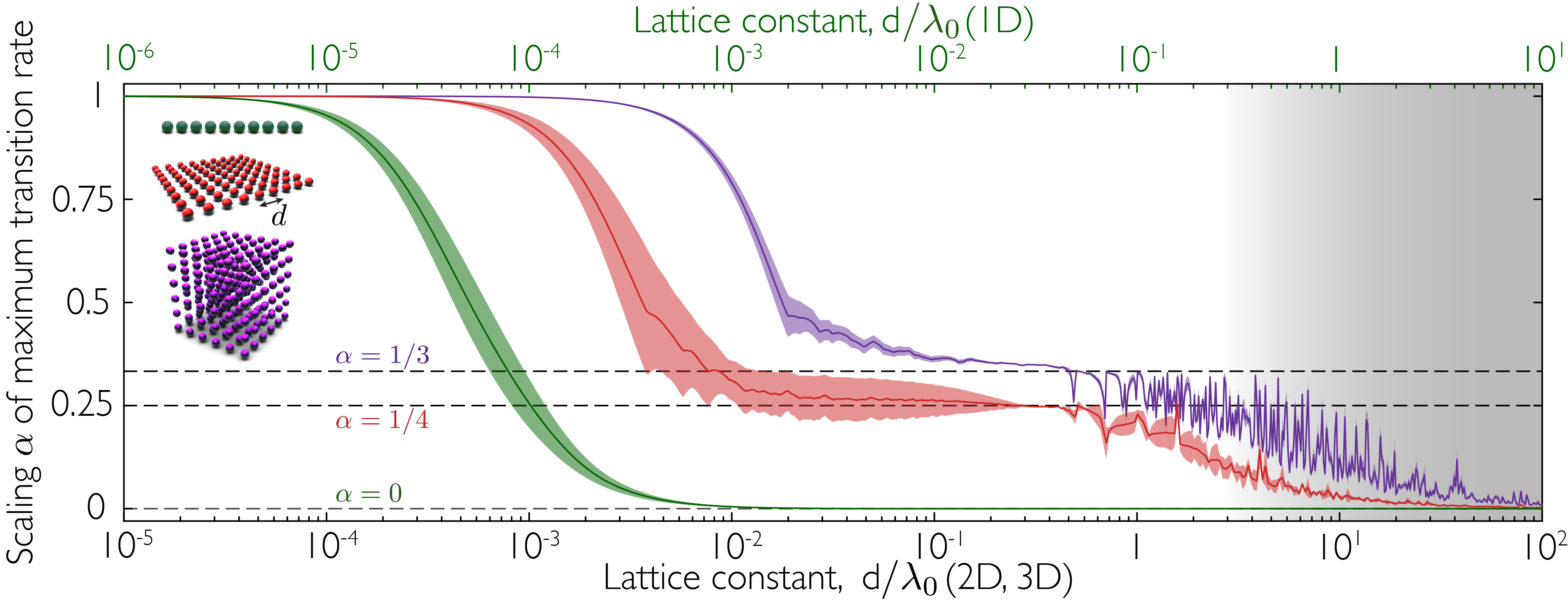}
             \caption{Finite size effects in the scaling of the largest transition rate (obtained from a best fit to $\Gamma_\text{max}=\beta N^\alpha \Gamma_0$) as a function of lattice constant. The atoms form a square lattice and are polarized parallel to one axis of the array. The fits are done over a region $N_\text{1D}\in [2, N^\text{max}_\text{1D}]$, where $N^\text{max}_\text{1D}=\{30000, 250, 40\}$ for 1D, 2D, and 3D, respectively. The colored regions represent the 1$\sigma$ confidence interval.  The gray area shows the region where the fit is not accurate (R-squared $<0.95$, see Sec.~\ref{app:scaling_Gamma_max} of SI). The upper axis shows the lattice constant exclusively for 1D arrays. Dashed lines represent the analytical scaling.}\label{fig3}
\end{figure*}

For large $N$, the numerical approximations to $R_\star$ given by the SDP follow the analytical scaling law of Eq.~\eqref{eq:array_scaling_law}.  For visualization purposes, we shift the data set corresponding to each lattice dimension by a multiplicative factor (a constant upward shift in logarithmic scale). These shifts do not affect the scaling and highlight the excellent agreement between the numerical results and the analytical scaling laws. \new{Since the dominant decay is delocalized, Fig.~\ref{fig2} validates our scaling law for $R_\star$.} Our results suggest that the SDP can be a valuable tool to obtain empirical scaling laws of $R_\star$ at large system sizes for more complicated systems that are analytically intractable.  

For sufficiently large atom numbers the scalings hold regardless of the lattice constant. For finite $N$, however, they depend on other parameters such as the lattice constant and lateral size $L=N^{1/\text{D}}d$, as shown in Fig.~\ref{fig3}. As we discuss in \new{Sec.~\ref{app:scaling_Gamma_max} of} the SI, by taking the limits of the expression for $\mathbf{\Gamma}$ in the appropriate order ($k_0d\rightarrow 0$ before $N\rightarrow\infty)$ we confirm that for $L\ll\lambda_0$, one recovers Dicke's scaling, i.e., $\Gamma_\text{max}=N\Gamma_0$. For arrays with large lattice constant, $d\gg\lambda_0$, following a similar procedure yields the limit of non-interacting atoms, i.e., $\Gamma_\text{max}\simeq\Gamma_0$. We determine the crossover between ``non-interacting'' and ``collective'' behaviors by identifying the parameters for which there is an asymptotic change in the scaling of $R_\star$, from linear to superlinear. For 2D and 3D arrays the number of atoms required for such crossover is $N^\text{(crit)}\simeq \eta (k_0 d)^6$, where $\eta=0.02$ and $5$, respectively. As expected, for large inter-particle distances the number of atoms required for $R_\star$ to be superlinear grows rapidly.

\textbf{Experimental implications.} 
\new{Here, we discuss how the scaling laws impact a broad range of areas, including quantum optics, out-of-equilibrium phase transitions, quantum simulation and fault-tolerant quantum computation.}

\new{\textit{(1) Transient superradiance beyond the Dicke limit.} Our findings on the scaling of $R_\star$ crucially address fundamental problems in quantum optics, such as transient and steady-state superradiance in extended systems~\cite{dicke1954coherence,Gross1982,Masson2022NatComm}. The scaling law sets a rigorous upper bound on the scaling of the superradiant burst. While this upper bound may be violated if light is collected only over a small solid angle, new scaling laws can be derived taking into account the detector aperture. Furthermore, in Sec.~\ref{app:g2} of the SI, we reveal  a connection between the early-time correlations that determine the appearance of a superradiant burst~\cite{Masson2022NatComm} (achieved under dynamical evolution) and $R_\star$ (which may be inaccessible by dynamics). While our approach does not capture dynamical evolution, it allows us to predict that the timescale of a superradiant burst from an initially fully excited D-dimensional array of atoms in free space scales as $\Gamma_0T_R\sim \beta^{-1}N^{(1-\text{D})/2\text{D}}\log(\beta N^{(D-1)/2}/2)$ (Sec.~\ref{app:Burst_Time} of SI). This timescale imposes constraints on the array size required for retardation effects to be negligible, thereby ensuring the Markovian evolution assumed in Eq.~(\ref{eq:lindbladME}).}

\begin{figure}[!t]
	\includegraphics[width=\columnwidth]{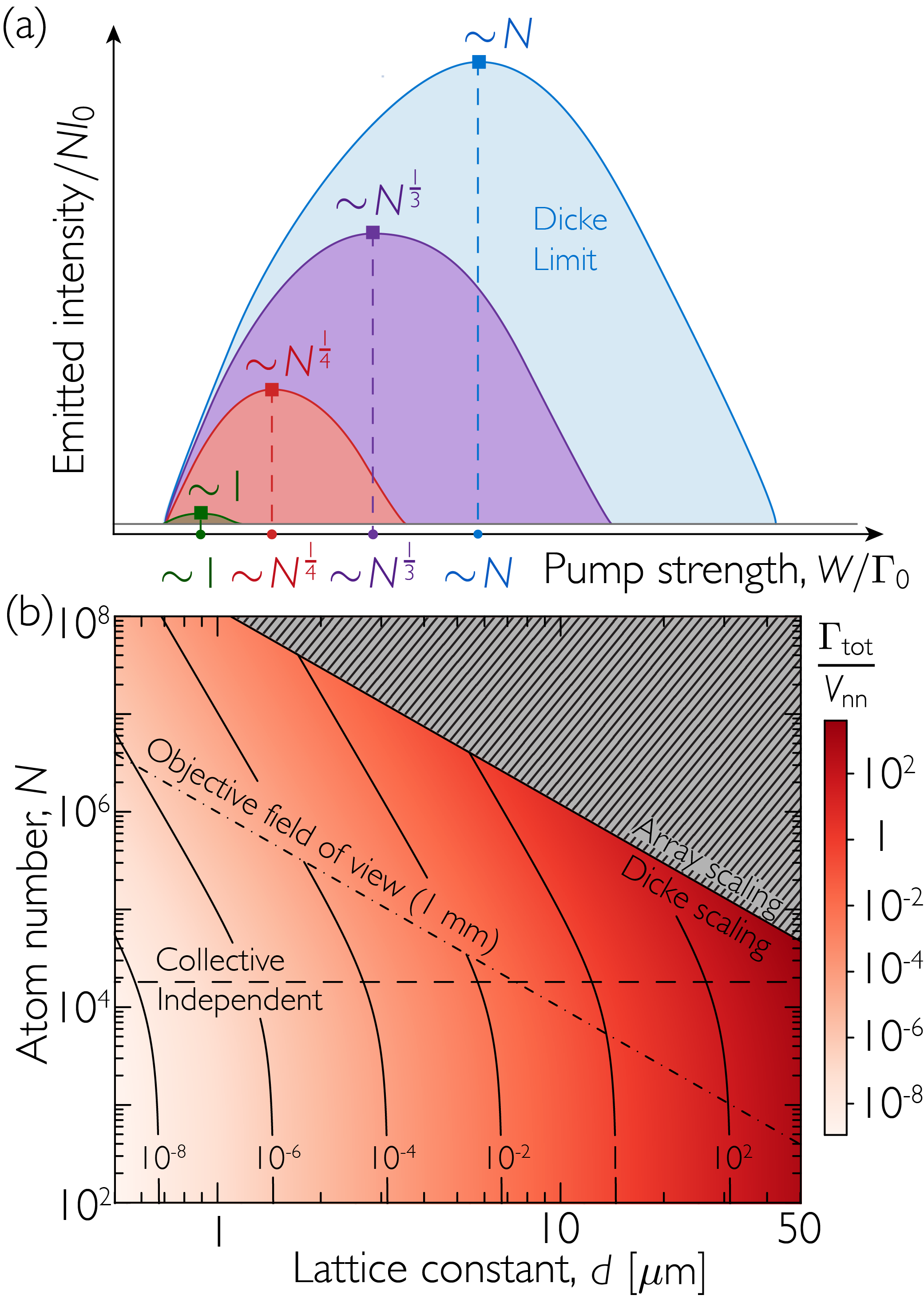}
	\caption{\new{Implications of our scaling laws for superradiant lasing in free space and quantum computing and simulation with Rydberg atom arrays. (a) Schematics of the expected scaling for the superradiant lasing region for incoherently pumped atoms in 1D (green), 2D (red), 3D (purple) arrays and in a cavity (blue). (b) Ratio between the largest total decay $\Gamma_\text{tot}$ (both correlated and independent) and the Rydberg-Rydberg interaction $V_\text{nn}$, for the $53S_{1/2}$ state in 2D Rydberg arrays of $^{87}\text{Rb}$ atoms. Collective decay for the $53S_{1/2}\to52P_{3/2}$ transition exceeds the total independent decay above the dashed line, following Dicke-like scaling since $N^{1/\text{D}}d < \lambda_0$. The dashed-dotted line indicates the state-of-the-art Rydberg arrays limit given by the objective field of view ($\sim 1 \, \text{mm}$ size). The region (striped) where the many-body decay rate of the transition follows the array scaling of Eq.~\eqref{eq:array_scaling_law} has been excluded for simplicity.}}\label{fig:Applications}
\end{figure}

\new{\textit{(2) Superradiant lasing in free space.} Superradiant lasing, where incoherently pumped atoms spontaneously radiate coherent light, is known to occur in cavities~\cite{Meiser2009, bohnet}. It is, however, unknown whether this phenomenon can occur in other environments. By means of our scaling laws, in Sec.~\ref{sec:SR_Laser} of the SI, we derive an upper bound on the emitted intensity $I_\star\sim \hbar \omega_0 R_\star$, which occurs at the optimal pump rate $W_\star = 2R_\star/N$. This upper bound is tight for the Dicke limit (i.e., for atoms in a cavity), yielding $W_\star = N\Gamma_0/2$. Our result for $I_\star$ rules out superradiant lasing for 1D arrays in free space (as there is never a superlinear scaling for the intensity). It suggests that a superradiant phase transition might occur for 2D and 3D arrays, with a scaling of the lasing region determined by Eq.~(\ref{eq:array_scaling_law}) [Fig.~\ref{fig:Applications}(a)].}

\new{\textit{(3) Driven-dissipative Dicke phase transition.} Coherently pumped atoms in a cavity exhibit a second order phase transition (at a critical pumping rate $N\Gamma_0/2$) from a magnetized phase characterized by a collective polarization to a paramagnetic phase~\cite{Hepp1973, narducci,carmichael_1980,Ferioli2023nonequilibrium}. Beyond the well-studied case of atoms in a cavity, this phenomenon -- also called collective resonance fluorescence --  has remained largely unexplored until recently~\cite{Ferioli2023nonequilibrium,Agarwal2024,Goncalves2024,Ostermann2024,Ruostekoski2024}. Our scaling laws can be harnessed to readily show that $R_\star$ sets the scaling of both the threshold drive intensity, $\eta_c\sim \sqrt{\Gamma_0 R_\star/N}$, and of the emitted light intensity below threshold $I\sim \hbar\omega_0 R_\star$ (see Sec.~\ref{sec:Coherent_Pumping} of SI). These analytical results are in agreement with recent numerical studies on collective resonance fluorescence of free space arrays~\cite{Ostermann2024,Ruostekoski2024}.}

\new{\textit{(4) Quantum error correction and typical states.} The scaling law~\eqref{eq:array_scaling_law} indicates that correlated decay may hamper quantum error correction~\cite{Aharonov2006fault,Preskill2013sufficient,lemberger2017effect}, as the error rate per qubit scales (in the worst case) as $\sim R_\star/N$, which grows with $N$ in 2D and above. Nevertheless, in Sec.~\ref{app:typicalstates} of SI, we prove that the decay rate for typical (Haar-random) states is close to $N\Gamma_0/2$. Therefore, most states in the Hilbert space do not experience correlated decay, due to random phases between the qubits. This does not mean that the scaling laws for $R_\star$ are irrelevant in practice, since even simple states like the product state $\ket{+}^{\otimes N}$ may be superradiant.}

\new{\textit{(5) Quantum simulators and processors with Rydberg atoms.} For microwave transitions, such as between Rydberg states (where $\lambda_0 \approx 10 \, \text{mm}$), typical experimental implementations lie within the collective decay regime. Indeed, collectively enhanced decay in a dense Rydberg ensemble has been linked to Dicke-like superradiance~\cite{Wang2007}. This may increase leakage error rates out of the computational subspace in quantum processors~\cite{cong}, and reduce the coherence times of quantum simulators.}

\new{As a demonstration, we consider the setup of Ref.~\cite{Evered2023}, which uses a 2D array of tweezer-trapped $^{87}\text{Rb}$ atoms. The same platform is also used to simulate complex many-body quantum dynamics~\cite{Manovitz2025}.  Although qubits are encoded in long-lived hyperfine ground states, the atoms are excited to Rydberg states during the entangling gate operations. To implement entangling gates in parallel across the entire array, Rydberg states are macroscopically populated, which can potentially lead to collectively enhanced decay. We compute the ratio between the total decay rate $\Gamma_\text{tot}$ from the $53S_{1/2}$ Rydberg level (including collective decay) and the coherent Rydberg-Rydberg interaction rate $V_{nn}=C_6/d^6$ between neighboring atoms [Fig.~\ref{fig:Applications}(b)]. For simplicity, we plot $\Gamma_{\text{tot}}/V_{\text{nn}}$ in the region where $\sqrt{N} d < \lambda_0$, such that the contribution of collective decay to $\Gamma_{\text{tot}}$ has a Dicke-like scaling (corresponding to the leftmost plateau in Fig.~\ref{fig3}). For reference, the linear size of state-of-the-art Rydberg arrays, which is on the order of 1 \text{mm} and primarily limited by the objective field of view~\cite{endres2024}, is also indicated on the plot. We define the total decay rate as $\Gamma_\text{tot} \equiv (N-2)\Gamma_{53S_{1/2}\rightarrow 52P_{3/2}}/4 + \Gamma_\text{sp} + \Gamma_\text{bbr}$, where the first term represents the dominant collective decay from the Rydberg level $53S_{1/2}$ to the level $52P_{3/2}$ (in the Dicke limit), while the latter two terms denote the contributions of all possible spontaneous ($\Gamma_\text{sp}$) and black-body induced ($\Gamma_\text{bbr}$) transitions from the Rydberg level at room temperature (see Sec.~\ref{app:Rydberg_Collective} of SI for a precise definition).  The decay-induced gate error, approximately $\Gamma_{\text{tot}}/\Omega$, can then be estimated by rescaling the ratio in Fig.~\ref{fig:Applications}(b) by $V_{nn}/\Omega$, where $\Omega$ is the Rabi frequency used to realize the gate.}

\new{While collective decay contributes negligibly to the gate error for current experiments with a few hundred atoms, we estimate that for system sizes of a few thousand atoms it might become significant (see Sec.~\ref{app:Rydberg_Collective} of SI). We expect such effects to be observable in state-of-the-art experimental platforms, which have achieved $N \sim 6000$ atoms~\cite{endres2024}. The results in Fig.~\ref{fig:Applications}(b) for $2\text{D}$ arrays also hold for 3D arrays, as the ratio is nearly independent of dimensionality in the Dicke regime considered here. This suggests that decoherence due to collective decay becomes more relevant in 3D arrays, where larger arrays can be built. Our analysis indicates that further investigation of correlated decay in large-scale Rydberg atom arrays is critical to assess the scalability of these platforms for quantum computation and simulation purposes.}

\textbf{Outlook.} 
\new{Our results highlight a new class of universal scaling laws that govern the fundamental limits of decay in many-body quantum systems. These findings open several avenues for both experimental exploration and theoretical development. One important direction is quantum metrology, where correlated decay may impose fundamental constraints.} Recent experiments on lattice clocks~\cite{Hutson2024} and spin squeezing~\cite{block2023universal, Bornet2023, eckner, franke} have investigated the role of Hamiltonian power-law dipole-dipole interactions. The dissipative counterpart of the interaction is typically neglected (as dephasing noise is currently the main source of error), although it sets a fundamental limit on the time available to generate and utilize metrologically-useful states. While the decay rate of typical (Haar-random) states is close to $N\Gamma_0/2$, \new{quantum metrology protocols often rely on a carefully selected set of states (e.g., Dicke states, squeezed states), which provide metrological advantage~\cite{Oszmaniec2016}. Investigating the many body decay of such states will prove essential to establish possible limitations on such schemes.}

From a theoretical perspective, our approach illustrates the power of quantum approximation techniques to predict relevant properties of open many-body systems. \new{Our formalism can be extended to yield upper bounds on the rates of change of higher-order observables such as $k$-point correlation functions (in this work, $\hat{n}_{\text{exc}}$ corresponds to $k = 1$). Replacing $\hat{n}_{\text{exc}}$ in Eq.~\eqref{eq:Rate} with a $k$-local observable, the ``auxiliary'' Hamiltonian $\hat{H}_\Gamma$ will in general be $(k+1)$-local, and the optimal product state provides an approximation with a multiplicative error of at most $3^{k+1}$, independent of system size~\cite{bravyi2019approximation}. Furthermore, our approach} could also be adapted to study the steady-state behavior of driven-dissipative systems, to predict the scaling of correlations functions or other physical observables~\cite{Wang2024}. More advanced SDP relaxations such as the quantum Lasserre hierarchy~\cite{parekh2021application} could yield tighter bounds for generic systems. These methods are not restricted to spin models, and could be extended to study ensembles of interacting fermions or bosons, as well as to disordered systems. Moreover, invoking time-reversal arguments, our scalings also apply to the maximal absorption rate, which may have implications for quantum batteries~\cite{campaioli2023} and light harvesting protocols~\cite{scholes-photo}. Given their generality, we anticipate that these ideas will become a powerful tool to investigate universal properties of large scale many-body open quantum systems.

\textbf{Acknowledgments.} We thank Antonio Acin,  Paul R. Berman, Darrick E. Chang, Tobias Haug, Simon B. J\"{a}ger and Leo Zhou for helpful discussions. We are grateful to Silvia Cardenas-Lopez for her help with the design of the first figure. We acknowledge support by the National Science Foundation through the CAREER Award (No. 2047380), the Air Force Office of Scientific Research through their Young Investigator Prize (grant No. 21RT0751), as well as by the David and Lucile Packard Foundation. JP acknowledges support from the National Science
Foundation (Award No. PHY-2317110). The Institute for Quantum Information and Matter is an NSF Physics Frontiers Center.

%% Begin Supplementary Information

\let\oldaddcontentsline\addcontentsline% Store \addcontentsline
\renewcommand{\addcontentsline}[3]{}% Make \addcontentsline a no-op
\bibliography{main}
\vspace{.5cm}
%\printbibliography[heading=none]
\let\addcontentsline\oldaddcontentsline% Restore \addcontentsline

\balancecolsandclearpage
%\newrefsection
\appendix

\newpage
\onecolumngrid

\begin{center}
\vspace{1cm}
\textbf{\large Supplementary Information}
\end{center}
\def\set@footnotewidth{\onecolumngrid}% <<<<<<<<<<<<<<<<
\def\footnoterule{\kern-6pt\hrule width 1.5in\kern6pt}%
\renewcommand{\appendixname}{}
% \\renewcommand{\thesection}{\Roman{section}}
\renewcommand{\thesubsection}{\MakeUppercase{\alph{section}}.\arabic{subsection}}
\renewcommand{\thesubsubsection}{\Alph{subsection}.\arabic{subsubsection}}
\makeatletter
\renewcommand{\p@subsection}{}
\renewcommand{\p@subsubsection}{}
\def\@hangfrom@section#1#2#3{\@hangfrom{#1#2}.\quad\MakeTextUppercase{#3}}%

\makeatother

\renewcommand{\thefigure}{S\arabic{figure}}
\setcounter{figure}{0}
\setcounter{secnumdepth}{3}
\renewcommand{\theHfigure}{figure.section.\thesection.\thefigure}

\vspace{-1cm}
\tableofcontents

\section{General theory}

\new{This section contains additional material on the general theory presented in the main text. The robustness of the scaling laws for $R_\star$ for the generalization of the master equation~(\ref{eq:lindbladME}) is studied in Sec.~\ref{sec:disorderandmore}. In Sec.~\ref{app:Bound_Observable}, we show that $R_\star$ sets a bound on the rate of change of generic observables. For systems where the dissipative interactions are in-phase, the corresponding scaling laws are derived in Sec.~\ref{app:inphase}. In Sec.~\ref{app:translation}, we prove the lower bound for delocalized decay given in Eq.~(\ref{eq:product_lower}). In Sec.~\ref{appendix:lb} we derive the lower bound for translationally invariant systems using a spin-wave ansatz for the collective jump operators. We show that the scaling law for delocalized system is not true for generic systems in Sec.~\ref{app:reductio}. Finally, the decay rate for typical Haar random states is derived in Sec.~\ref{app:typicalstates}.}

\subsection{Robustness of the scaling laws to disorder, arbitrary local Hamiltonian and dissipative terms}\label{sec:disorderandmore}
Here, we show that the scaling law $R_\star \sim N\Gamma_\text{max}$ holds much more generally than under the assumptions we considered in Eq.~\eqref{eq:lindbladME}.

{\bf Disorder.} We include disorder by considering a decoherence matrix 
\begin{equation}\label{eq:Gamma_prime_disorder}
    \mathbf{\Gamma}^{\prime} = \mathbf{\Gamma} + \mathbf{\Gamma}_{\text{disorder}},
\end{equation}
with the disorder matrix $\mathbf{\Gamma}_{\text{disorder}}$ Hermitian but not positive semidefinite in general. For the decoherence matrix to be physically valid, we demand that $\mathbf{\Gamma}^\prime \succeq 0$. Weyl's inequality~\citesupp[p.~239]{Horn_Johnson_2012supp} yields
\begin{equation}
    \left| R_\star^\prime - R_\star \right| \leq \norm{\hat{H}_{\Gamma_{\text{disorder}}}}
\end{equation}
and
\begin{equation}
    \left| \Gamma^{\prime}_{\text{max}} - \Gamma_{\text{max}} \right| \leq \norm{\mathbf{\Gamma}_{\text{disorder}}},
\label{eq:disorderweyl}
\end{equation}
where $R^\prime_*$ and $\Gamma^\prime_{\text{max}}$ are the new maximal decay rate and largest transition rate respectively. Here, $\hat{H}_{\Gamma_{\text{disorder}}}$ is defined analogously to Eq.~\eqref{eq:haux} with the decoherence matrix $\mathbf{\Gamma_{\text{disorder}}}$. \new{We now provide a sufficient condition on the disorder for the scaling law to be robust. In general, we can write
\begin{equation}
    \mathbf{\Gamma}_{\text{disorder}} = \mathbf{\Gamma}_{\text{disorder}}^{(+)} + \mathbf{\Gamma}_{\text{disorder}}^{(-)},
\end{equation}
where $\mathbf{\Gamma}_{\text{disorder}}^{(\pm)}$ is the projection of $\mathbf{\Gamma}_{\text{disorder}}$ onto its eigenspace with positive (negative) eigenvalues. Applying the upper bound~\eqref{eq:master_bound} and the triangle inequality to Eq.~\eqref{eq:disorderweyl}, we have
\begin{equation}
    \left| R_\star^\prime - R_\star \right| \leq O\left(N \norm{\mathbf{\Gamma}_\text{disorder}^{(+)}} + N \norm{\mathbf{\Gamma}_\text{disorder}^{(-)}}\right) \leq O\left(N \norm{\mathbf{\Gamma}_{\text{disorder}}}\right).
\end{equation}
Therefore, the scaling law for $R_\star$ is robust if $\norm{\mathbf{\Gamma}_{\text{disorder}}}/\Gamma_{\text{max}} \ll 1$. Since $\mathbf{\Gamma}_{\text{disorder}}$ is an $N \times N$ matrix, this robustness condition can be efficiently verified numerically (or analytically, for certain types of disorder). }

{\bf Local Hamiltonian and dissipation.} We consider a general $k$-local qubit Hamiltonian $H^{(k)}$ which can be written as a linear combination of $Q$ Pauli strings of weight $k$
\begin{equation}
    \hat{H}^{(k)} = \hbar\sum_{j=1}^{Q} \eta_j \hat{P}_j^{(k)}.
\end{equation}
Each $\hat{P}_j^{(k)}$ is a tensor product of $k$ single-qubit (non-identity) Pauli operators acting on $k$ qubits, and act identically on the remaining $N-k$ qubits. The coefficients $\eta_j \in \mathbb{R}$ are constants assumed to be independent of $N$, i.e., proportional to $\Gamma_0$. On top of that, we can also add $Q$ independent $k$-local dissipation channels denoted by Lindblad dissipators of the form 
\begin{equation}
    \sum_{j=1}^{Q} \kappa_j (\hat{P}_j^{(k)} \hat{\rho} \hat{P}_j^{(k)} - \hat{\rho}), 
\end{equation}
where we have used the fact that the Pauli strings $\hat{P}_j^{(k)}$ satisfy $\hat{P}_j^{(k) \dag} = \hat{P}_j^{(k)}$ and $\hat{P}_j^{(k) 2} = 1$ to simplify the expression. The rates $\kappa_j$ are positive coefficients that we assume to be independent of $N$, \new{i.e., proportional to $\Gamma_0$}. Physically, $\kappa_j$ represents the decay rate of the local dissipative channel with jump operator $\hat{P}_j^{(k)}$. Adding these terms modify the master equation~\eqref{eq:lindbladME} to
\begin{equation}
    \dot{\hat{\rho}} = -\frac{i}{\hbar}[\hat{H} + \hat{H}^{(k)},\hat{\rho}] + \sum_{i,j=1}^{N} \Gamma_{ij} \left(\sm_i \hat{\rho} \sp_j - \frac{1}{2}\{\sp_j \sm_i, \hat{\rho} \} \right) + \sum_{j=1}^{Q} \kappa_j (\hat{P}_j^{ (k)} \hat{\rho} \hat{P}_j^{(k)} - \hat{\rho}).
\end{equation}
\new{Correspondingly, the auxiliary Hamiltonian $\hat{H}_\Gamma$ has the general form
\begin{equation}
    \hat{H}_\Gamma = \sum_{i,j=1}^{N} \Gamma_{ji} \sp_i \sm_j - i \sum_{j=1}^{Q} \eta_j [\hat{P}_j^{(k)},\hat{n}_{\text{exc}}] - \sum_{j=1}^{Q} \kappa_j \left(\hat{P}_j^{(k)} \hat{n}_{\text{exc}} \hat{P}_j^{(k)} - \hat{n}_{\text{exc}}\right).
\end{equation}}

We now bound the contributions of the additional Hamiltonian and dissipative terms to the decay rate $R$, defined as the expectation value of $\hat{H}_\Gamma$. First, let us examine the contribution of the local Hamiltonian $\hat{H}^{(k)}$. From the expression $\hat{n}_{\text{exc}} = \sum_{i=1}^{N} (\mathbf{I}_2+\hat{\sigma}_i^z)/2$,
\begin{equation}
    R = -\frac{i}{2} \sum_{j=1}^{Q} \eta_j \braket{[\hat{P}_j^{(k)},\sum_{i=1}^{N} \hat{\sigma}_i^z]}.
\end{equation}
The commutator term produces a linear combination of at most $k$ Pauli strings, since each $\hat{P}_j^{(k)}$ is supported only on $k$ qubits. Using the fact that the expectation value of a Pauli string has a magnitude of at most $1$, we obtain the bound
\begin{equation}
    R \leq \text{const.} \times k \sum_{j=1}^{Q} |\eta_j| = O(kQ \Gamma_0). 
\end{equation}
Next, we examine the contribution from the local dissipation:
\begin{equation}
    R = \sum_{j=1}^{Q} \kappa_j \left(\braket{\hat{n}_{\text{exc}}} - \braket{\hat{P}_j^{(k)} \hat{n}_{\text{exc}} \hat{P}_j^{(k)}} \right) = \sum_{j=1}^{Q} \frac{\kappa_j}{2} \sum_{i=1}^{N} \left( \braket{\hat{\sigma}_i^z} - \braket{\hat{P}_j^{(k)}\hat{\sigma}_i^z \hat{P}_j^{(k)} }\right).
\end{equation}
The term in the parenthesis has a non-vanishing contribution if qubit $i$ lies within the support of $\hat{P}_j^{(k)}$. A similar argument as before gives the bound
\begin{equation}
    R \leq \text{const.} \times k \sum_{j=1}^{Q} \kappa_j = O(kQ\Gamma_0).
\end{equation}
The upshot is that adding the local Hamiltonian and dissipative terms can only contribute an $O(kQ\Gamma_0)$ correction to the decay rate and thus
\begin{equation}\label{eq:R_star_locHamandDiss}
    R_\star \sim N\Gamma_{\text{max}} + O(kQ\Gamma_0).
\end{equation}
The size of this correction depends on $k$ and $Q$. Of course, if we include all possible choices of $k$-local Pauli strings, $Q = 3^k \binom{N}{k}$ dominates over the scaling of $N\Gamma_{\text{max}}$ for $k \geq 2$. However, a vast majority of such Pauli strings contain spatially non-local interactions. We impose geometric locality on $\hat{P}_j^{(k)}$ by demanding that each of the $k$ qubits involved are located within a $D$-dimensional ball of constant radius $r$. For an ordered system in $D$ spatial dimensions, we have $Q = O(\text{const.}^k \times N r^{kD}/k!)$, where the constant comes from the $3^k$ choices of Pauli operators and also other geometrical factors (such as the coordination number). As long as the locality $k$ of the interaction is independent of the system size, the correction to $R_\star$ is only linear in $N$, which does not affect our scaling laws. We remark that a lattice geometry is not strictly necessary here, and a similar counting argument for $Q$ can be made for a spatially disordered system of qubits as long as the interactions described by $\hat{P}_j^{(k)}$ remain spatially local.

{\bf Driven-dissipative systems.}
One can generalize Eq.~\eqref{eq:lindbladME} to include both coherent and incoherent drives, described by the master equation 
\begin{equation}\label{eq:lindblad_ME_incoher_drive}
\begin{split}
    \dot{\hat{\rho}} &= -i\left[\hat{H} + \sum_j (\eta_j \spl_j+\text{h.c.}),\hat{\rho}\right] + \sum_{i,j=1}^{N} \Gamma_{ij} \left(\sm_i \hat{\rho} \sp_j - \frac{1}{2}\{\sp_j \sm_i, \hat{\rho} \} \right) + \sum_{i,j=1}^{N} W_{ij} \left(\sp_i \hat{\rho} \sm_j - \frac{1}{2}\{\sm_j \sp_i, \hat{\rho} \} \right),
\end{split}
\end{equation}
where the coherent and incoherent driving strengths are given by the constants $\eta_j$ and $W_{ij}$, respectively. The incoherent terms can arise from both an external pump or a finite temperature reservoir, allowing for the possibility of collective pumping. The instantaneous correlated many-body decay rate in Eq.~(\ref{eq:Rate}) now becomes $R=\avg{\hat{H}_\Gamma} + \avg{V_W} + \avg{V_\eta}$ where
\begin{equation}\label{eq:Haux_incoher_dr}
    V_W =  -\sum_{i,j}W_{ij}\spl_i\smi_j +\sum_{j=1}^N W_{jj}\hat{\sigma}^z_j,\; \;\text{and}\;\,
    V_{\eta} =i\sum_{j=1}^{N} (\eta_j \sp_j - \text{h.c.}).
\end{equation}

Assuming non-extensive parameters, both $V_{\eta}$ and the second term in $V_W$ can only shift $R_\star$ by at most $O(N)$ [see the discussion after Eq.~(\ref{eq:R_star_locHamandDiss})], which does not affect the scaling of $R_\star$. Thus, it suffices to consider the shifted decoherence matrix $\mathbf{\Gamma} - \mathbf{W}$ arising from incoherent pumping, where $\mathbf{W} = (W_{ij})_{i,j=1}^{N}$, as $R_\star$ is asymptotically equivalent to the largest eigenvalue of $\hat{H}_{\Gamma-W}$. One can thus identify the pumping with an effective disorder, and associate $\mathbf{\Gamma}_{\text{disorder}} \equiv -\mathbf{W}$, which implies that the scaling of $R_\star$ is unaffected if $\norm{V_{W}} < R_\star$. Similarly, the scaling of $\Gamma_{\text{max}}$ is unaffected if $\norm{\mathbf{W}} < \Gamma_{\text{max}}$. 

For a finite temperature reservoir, $\mathbf{\Gamma} - \mathbf{W}$ is independent of the mean number of thermal bosons $\bar{n}_{\text{th}}$ in the reservoir~\citesupp{Carmichael1993opensupp}. Thus, assuming that $\bar{n}_{\text{th}}$ does not scale with $N$ such that the second term in $V_W$ only contributes at most $O(N)$ to $R_\star$, the scaling of $R_\star$ is equivalent to that of a zero-temperature bath. This indicates that our scaling laws obtained from the analysis of Eq.~\eqref{eq:lindbladME} remain valid even at finite temperatures and, more generally, for driven-dissipative systems, except for very strong driving strengths where $\norm{V_W}$ or $\norm{V_\eta}$ scale faster than $R_\star/N$.

\new{\subsection{Bounds on rates of change for general observables}\label{app:Bound_Observable}
Although $R_\star$ is defined as the maximal decay rate of the many-body excitation in the system, here we show that it can be used to obtain upper bounds on the rates of change for general observables. As discussed in Sec.~\ref{sec:disorderandmore}, the dominant contribution arises from the collective decay. Thus, it suffices to consider the simplified model with only collective decay (to leading order in $N$).}

\new{To begin, let us consider an arbitrary positive operator $\hat{A}$. Assuming $\lVert\hat{H}\rVert \ll R_\star$, the rate of change is given by
\begin{equation}
    -\frac{d}{dt}\braket{\hat{A}} = \frac{1}{2} \left(\braket{\hat{H}_{\Gamma} \hat{A}} + \braket{\hat{A} \hat{H}_{\Gamma}} \right) - \sum_{\mu=1}^{N}\Gamma_\mu \braket{\hat{c}_\mu^\dag \hat{A} \hat{c}_\mu}.
\end{equation}
Now, we have the bounds
\begin{equation}
\begin{aligned}
    \braket{\hat{H}_\Gamma \hat{A}} &= \Tr\left(\hat{H}_\Gamma \hat{A} \hat{\rho}\right) \\
    &\leq \norm{\hat{H}_\Gamma} \norm{\hat{A}} \norm{\hat{\rho}}_1 & & \text{(H\"older inequality)} \\
    &= R_\star \norm{\hat{A}},
\end{aligned}
\end{equation}
and
\begin{equation}
    \sum_{\mu=1}^{N}\Gamma_\mu \braket{\hat{c}_\mu^\dag \hat{A}\hat{c}_\mu} \geq 0.
\end{equation}
Therefore, for bounded positive operators, the decay rate is bounded by $R_\star$ via
\begin{equation}
    -\frac{d}{dt}\braket{\hat{A}} \leq R_\star \norm{\hat{A}}.
\label{eq:positiveop_Rstarbound}
\end{equation}
Physically, $\braket{\hat{A}}$ can represent the outcome probability of a quantum measurement, such as the fidelity against a target subspace. }

\new{Another relevant class of observables is an arbitrary Hermitian operator constructed from the sum of $k$-local bounded Hermitian operators, i.e.,
\begin{equation}
    \hat{B} = \sum_{i=1}^Q \hat{b}_i, \quad \norm{\hat{b}_i}\leq 1,
\end{equation}
where each operator $\hat{b}_i$ acts non-trivially on at most $k$ qubits. Note that $\lVert\hat{B}\rVert$ can scale extensively with $N$. As a physical example, $\hat{B}$ can represent $k$-body correlation functions. We now seek a bound for the maximum rate of change in terms of $R_\star$:
\begin{equation}
\begin{aligned}
    \left|\frac{d}{dt}\braket{\hat{B}}\right| &\leq \sum_{\mu=1}^{N} \Gamma_\mu \left|\braket{\hat{c}_\mu^\dag[\hat{B},\hat{c}_\mu]}\right| = \sum_{\mu=1}^{N} \Gamma_\mu \left| \Tr \left(\hat{c}_\mu^\dag [\hat{B},\hat{c}_\mu]\rho\right)\right| \\
    &= \sum_{\mu=1}^{N} \Gamma_\mu \left| \Tr \left(\rho^{1/2} \hat{c}_\mu^\dag [\hat{B},\hat{c}_\mu]\rho^{1/2} \right)\right| & & \text{(cyclic property of trace)} \\
    &\leq \sum_{\mu=1}^{N} \Gamma_{\mu} \norm{\rho^{1/2} \hat{c}_\mu^\dag}_F \norm{\rho^{1/2}}_F \norm{[\hat{B},\hat{c}_\mu]} & & \text{(H\"older inequality)} \\
    &= \sum_{\mu=1}^{N} \Gamma_\mu \sqrt{\braket{\hat{c}_\mu^\dag \hat{c}_\mu}} \norm{[\hat{B},\hat{c}_\mu]}.
\end{aligned}
\end{equation}
Here, $\lVert\cdot\rVert_F$ denotes the Frobenius norm. Since $\hat{B}$ and $\hat{c}_\mu$ are constructed from local bounded operators, we expect $\lVert[\hat{B},\hat{c}_\mu]\rVert$ to be bounded. For concreteness, let us assume the system is translationally invariant, such that
\begin{equation}\label{eq:Local_Observables}
    \norm{[\hat{B},\hat{c}_\mu]} \leq \frac{1}{\sqrt{N}} \sum_{i=1}^{Q} \sum_{j=1}^{N} \norm{[\hat{b}_i,\hat{\sigma}_j^-]}.
\end{equation}
Now, $\lVert[\hat{b}_j,\sigma_i^-]\rVert \leq 2$ (by the triangle inequality) if qubit $i$ is in the spatial support of the $k$-local operator $\hat{b}_j$, and $0$ otherwise. Therefore, we have $\lVert[\hat{B},\hat{c}_\mu]\rVert \leq 2kQ/\sqrt{N}$, and
\begin{equation}
\begin{aligned}
     \left|\frac{d}{dt}\braket{\hat{B}}\right| &\leq \frac{2kQ}{\sqrt{N}} \sum_{\mu=1}^{N} \Gamma_\mu \sqrt{\braket{\hat{c}_\mu^\dag \hat{c}_\mu}} \\
     &\leq \frac{2kQ}{\sqrt{N}} \sqrt{\sum_{\mu=1}^{N} \Gamma_\mu} \times \sqrt{\sum_{\mu=1}^{N} \Gamma_\mu \braket{\hat{c}_\mu^\dag \hat{c}_\mu}} & & \text{(Jensen's inequality)} \\
     &= 2kQ \sqrt{\Gamma_0} \times \sqrt{\braket{\hat{H}_\Gamma}} & & \left(\sum_{\mu} \Gamma_\mu = N\Gamma_0\right) \\
     &\leq 2kQ \sqrt{\Gamma_0 R_\star}.
\end{aligned}
\label{eq:localop_Rstarbound}
\end{equation}
Equations~\eqref{eq:positiveop_Rstarbound} and~\eqref{eq:localop_Rstarbound} are complementary bounds, which are useful for different classes of general observables. For specific observables, it may be possible to obtain tighter bounds with a better scaling in $N$. For example, if we choose $\hat{B} = \hat{n}_{\text{exc}}$, Eq.~\eqref{eq:localop_Rstarbound} yields $R \leq 2N\sqrt{\Gamma_0 R_\star}$, which is larger than the trivial upper bound $R \leq R_\star$ by a factor of $2N\sqrt{\Gamma_0/R_\star}$.}

\new{The rate of change of generic observables could also be affected by terms in Eq.~(\ref{eq:lindbladME}) that do not affect $R_\star$. We now discuss one such example.}

\new{{\bf Effects of dephasing.} In the following, we consider the case of collective dephasing of the system. The system evolution under pure dephasing has the form
\begin{equation}
    \Dot{\rho} = \sum_{ij} \kappa_{ij} \Big(\hat{\sigma}_j^z \hat{\rho}\hat{\sigma}_i^z - \frac{1}{2}\{\hat{\sigma}_i^z\hat{\sigma}_j^z,\hat{\rho}\}\Big) = \sum_{\mu} \kappa_{\mu} \Big(\hat{z}_\mu \hat{\rho}\hat{z}_\mu - \frac{1}{2}\{\hat{z}_\mu^2,\hat{\rho}\}\Big),
\end{equation}
where $\kappa_{ij}$ is the collective dephasing matrix, $\kappa_{\mu}\geq 0$ its eigenvalues and $\hat{z}_\mu \equiv \sum_j \beta_j^{(\mu)}\hat{\sigma}_j^z$ the collective jump operators.  $\vec{\beta}^{(\mu)}$ is a normalized eigenvector of $\boldsymbol{\kappa}$.
Dephasing does not affect $R_\star$ as it does not change the total number of excitations. It does, however, contribute to the dissipative dynamics of general observables.
In particular, for observables of the form Eq.~(\ref{eq:Local_Observables}), we can apply the same procedure described above and obtain
\begin{equation}
    \bigg\vert\frac{\text{d}}{\text{d}t}\langle \hat{B}\rangle\bigg\vert \leq 2kQ\sqrt{\kappa_0}\sqrt{\bigg\langle\sum_{ij}\kappa_{ij}\hat{\sigma}_i^z\hat{\sigma}_j^z\bigg\rangle},
\end{equation}
where $\kappa_0 = N^{-1} \sum_\mu \kappa_\mu$
This result can be combined with Eq.~(\ref{eq:localop_Rstarbound}) to obtain a general bound on the rate of change of local observables under collective dissipation and dephasing.}

\new{\subsection{Special case: In-phase interactions}\label{app:inphase}}

\new{In the special case where the dissipative interactions between all pairs of emitters are in-phase with the local decay, i.e., $\Gamma_{ij} \geq 0$, the scaling law for $R_\star$ can be easily derived. This includes, for example, the Dicke limit where $\Gamma_{ij} = \Gamma_0 > 0$.}

\new{In the non-trivial case where the sum of dissipative interactions, $S = \sum_{i\neq j} \Gamma_{ij}$, scales faster than $N$, the optimal product state is $\ket{+}^{\otimes N}$, where $\ket{+} = (\ket{g} + \ket{e})/\sqrt{2}$ denotes the uniform superposition of the ground and excited state. The many-body decay rate of this state is simply $S/4$, to leading order in $N$. Invoking the variational principle, we have $R_\star \geq S/4$. Now, using the result in Ref.~\citesupp{bravyi2019approximationsupp}, we obtain the upper bound $R_\star \leq 3S/2$. Thus, we establish the scaling law
\begin{equation}
    R_\star \sim S.
\end{equation}
For instance, this recovers the scaling $R_\star \sim N^2$ in the Dicke limit. Note that this scaling is not valid for the free-space arrays studied in the main text, since the matrix elements $\Gamma_{ij}$ can take on either sign, depending on the distance between the emitters.}

\subsection{Proof of lower bound for delocalized decay}\label{app:translation}
By means of the variational principle and Eq.~\eqref{eq:haux} we find
\begin{equation}
    R_\star = \max_{\ket{\psi}} \sum_\mu \Gamma_\mu \new{\braket{\psi|\hat{c}_\mu^\dag \hat{c}_\mu|\psi}} \geq \Gamma_{\text{max}} \lVert \hat{c}_1^\dag \hat{c}_1 \rVert.
\end{equation}
Substituting the definition of the collective operators $\hat{c}_\mu$, we arrive at
\begin{equation}
\begin{split}
    R_\star &\geq \Gamma_{\text{max}} \bigg\Vert \sum_{i,j} \alpha_i^{(1)*} \alpha_j^{(1)} \hat{\sigma}_i^+ \hat{\sigma}_j^- \bigg\Vert = \Gamma_{\text{max}} \bigg\Vert \sum_{i,j} \left|\alpha_i^{(1)}\right| \left|\alpha_j^{(1)}\right| \tilde{\sigma}_i^+ \tilde{\sigma}_j^- \bigg\Vert = \Gamma_{\text{max}} \max_{\ket{\psi}} \sum_{i,j} \left|\alpha_i^{(1)}\right| \left|\alpha_j^{(1)}\right| \braket{\psi|\tilde{\sigma}_i^+ \tilde{\sigma}_j^-|\psi} \\&\geq \frac{\Gamma_{\text{max}}}{4} \left( \sum_i \left|\alpha_i^{(1)}\right| \right)^2,
\end{split}
\label{eq:rstar_LB}
\end{equation}
where we have absorbed the phases $\phi_i$ of $\alpha_i^{(1)}$ into the lowering operators $\tilde{\sigma}_i^- = e^{i\phi_i} \hat{\sigma}_i^-$. The last inequality is obtained by choosing the product state $\ket{\psi} = \bigotimes_i (\ket{g}+\exp(-i\phi_i)\ket{e})_i/\sqrt{2}$ such that $\braket{\psi|\tilde{\sigma}_i^+ \tilde{\sigma}_j^-|\psi}= 1/4$ \new{if $i\neq j$ and $\braket{\psi|\tilde{\sigma}_i^+ \tilde{\sigma}_i^-|\psi}= 1/2$}. The lower bound thus depends on how the collective operator $\hat{c}_1$ is spatially supported on the $N$ qubits, quantified by the variance of $|\alpha_i^{(1)}|$, which reads
\begin{equation}
\begin{split}
    \text{Var}(|\alpha^{(1)}|) &= \frac{1}{N} \sum_i \left(\left|\alpha_i^{(1)}\right| - \overline{\left|\alpha^{(1)}\right|}\right)^2 = \frac{1}{N} - \frac{1}{N^2} \left( \sum_i \left|\alpha_i^{(1)}\right| \right)^2.
\end{split}
\end{equation}
The second line of the equation is obtained by using $\sum_i |\alpha_i^{(1)}|^2=1$, and $\overline{|\alpha^{(1)}|}=\sum_i \left|\alpha_i^{(1)}\right|/N $. Defining the relative fluctuation of $|\alpha_i^{(1)}|$ as $\Delta = \sqrt{\text{Var}(\left|\alpha^{(1)}\right|)} / \overline{\left|\alpha^{(1)}\right|}$ yields Eq.~\eqref{eq:product_lower}. Explicitly,
\begin{equation}\label{eq:Delta}
    \Delta^2 = \frac{N}{\norm{\vec{\alpha}^{(1)}}_1^2} - 1,
\end{equation}
where $\vec{\alpha}^{(1)} = (\alpha_1^{(1)} \ldots \alpha_N^{(1)})^T$ is the dominant eigenvector of $\mathbf{\Gamma}$, and $\lVert\vec{\alpha}^{(1)}\rVert_p$ is the $L^p$ vector norm. Using the inequality $\lVert\vec{\alpha}^{(1)}\rVert_2 \leq \lVert\vec{\alpha}^{(1)}\rVert_1 \leq \sqrt{N} \lVert\vec{\alpha}^{(1)}\rVert_2$ and $\lVert\vec{\alpha}^{(1)}\rVert_2 = 1$ by normalization, we have the bound
\begin{equation}
    0 \leq \Delta \leq \sqrt{N-1}.
\end{equation}
The upper bound $\Delta = \sqrt{N-1}$ is satisfied for independent emitters. Intuitively, $\Delta$ quantifies the spatial uniformity of the brightest collective jump operator $\hat{c}_1$. Thus, we define the delocalized decay regime to be $\Delta = O(1)$. For translationally invariant systems, $|\alpha_i^{(\mu)}| = N^{-1/2}$ for all $i = 1,\ldots,N$ and $\mu = 1,\ldots,N$, such that $\Delta = 0$.

\new{\subsection{Lower bound with a physically motivated ansatz}\label{appendix:lb}}

\new{Here, we derive a lower bound using a physically-motivated ansatz (instead of a product state ansatz) for translationally invariant systems. For these systems, the auxiliary Hamiltonian Eq.~(\ref{eq:haux}) can be written as $\hat{H}_\Gamma = \sum_\kk \Gamma_{\kk} \hat{c}^\dag_{\kk}\hat{c}_{\kk}$, where $\Gamma_{\kk}$ are the eigenvalues of $\boldsymbol{\Gamma}$ and $\hat{c}_{\kk} \equiv N^{-1/2} \sum_{j=1}^N\exp(\ii \kk\cdot \rr_j)\smi_j$.The ansatz state is defined by applying the same jump operator $\hat{c}_{\kk}$ with wavevector $\kk$ $m$ times  to the fully excited state, i.e.,
\begin{equation}
    \ket{\Psi_{m,\kk}} = \frac{1}{\mathcal{N}_m}\big(\hat{c}_{\kk})^m\ket{e}^{\otimes N} 
   % \equiv \mathcal{N}_m^{-1}\ket{\psi_{m,\kk}},
\end{equation}
where the normalization constant reads~\citesupp{silvia2023manysupp}
\begin{equation}
    \mathcal{N}_m^{-2} = \frac{m!}{N^m}\binom{N}{m}.
\end{equation}
Finding the expectation value of the auxiliary Hamiltonian over this ansatz reduces to computing expectation values of the form 
\begin{equation}\label{eq:Step_1_ansatz}
\begin{split}
    \bra{\Psi_{m,\kk}}\hat{c}_{\qq} ^\dag \hat{c}_{\qq} \ket{\Psi_{m,\kk}} =& \frac{1}{\mathcal{N}_m}\frac{N^m}{m!}\sum_{\mathcal{J}_m,\mathcal{K}_m\subseteq [N]} \bra{\mathcal{J}_m}{\hat{c}_{\qq}^\dag \hat{c}_{\qq}}\ket{\mathcal{K}_m}e^{i\kk \cdot\big(\sum_{j\in \mathcal{J}_m}\mathbf{r}_{j}-\sum_{k\in\mathcal{K}_m}\mathbf{r}_{k}\big)},
\end{split}
\end{equation}
where $\mathcal{J}_m,\mathcal{K}_m$ are subsets of cardinality $m$ of the set $\cpare{1,\dots,N}\equiv[N]$, and their corresponding sums run over all possible subsets of fixed cardinality. The sum in Eq.~(\ref{eq:Step_1_ansatz}) thus runs over all possible different ways to distribute $N-m$ excitations between $N$ qubits. Accordingly, $\ket{\mathcal{J}_m}=\hat{\sigma}_{j_1}^-\dots\hat{\sigma}_{j_m}^-\ket{e}^{\otimes N}$, with $\cpare{j_1,\dots,j_m}=\mathcal{J}_m$, represents a configuration where qubit $i$ is in the ground state if $i\in\mathcal{J}_m$ and in the excited state otherwise. The expectation value $\bra{\mathcal{J}_m}{\hat{c}_{\qq}^\dag \hat{c}_{\qq}}\ket{\mathcal{K}_m}$ is non-zero only if $\ket{\mathcal{J}_m}$ and $\ket{\mathcal{K}_m}$ differ only by the position of one excitation. Hence, substituting the expressions for $\hat{c}_{\qq}^\dag$ and $\hat{c}_{\qq}$ we obtain
\begin{equation}
\begin{split}
    \bra{\Psi_{m,\kk}}\hat{c}_{\qq}^\dag \hat{c}_{\qq} \ket{\Psi_{m,\kk}} =& \frac{(N-m)(N-m-1)}{N(N-1)} + \frac{m(N-m)}{N(N-1)}f_N(\kk-\qq),
\end{split}
\end{equation}
where we have defined $f(\kk) \equiv \frac{1}{N}\sum_{i,j=1}^N e^{\ii \kk\cdot(\rr_i-\rr_j)}$.}

\new{We take $\kk=\kk_1$ to be the wavevector associated to the jump operator with largest transition rate $\Gamma_{\text{max}}$, and $\qq=\kk_\mu=\pi(\mu_x,\mu_y,\mu_z)/N_\text{1D} d$ where $\mu_{x,y,z}$ are integers, corresponding to the wavevectors of all different collective jump operators. In that case, $f(\kk_1-\kk_\mu)=N\delta_{1,\mu}$ and one readily finds
\begin{equation}\label{eq:gamma_m_Phys_ansatz}
\begin{split}
    \bra{\Psi_{m,\kk_1}} \hat{H}_\Gamma \ket{\Psi_{m,\kk_1}} &= \frac{(N-m)(N-m-1)}{N-1}\Gamma_0+\frac{m(N-m)}{N-1}\Gamma_\text{max}.
\end{split}
\end{equation}}

\new{To obtain a better lower bound on the largest eigenvalue of $\hat{H}_\Gamma$, one has to optimize over $m$, which we denote as $m_\star$. Two limiting cases can be immediately identified. For independent atoms (i.e., $\boldsymbol{\Gamma}=\Gamma_0\mathds{1}$) the largest decay is $\Gamma_\text{max}=\Gamma_0$ and thus $m_*=0$. For $\Gamma_\text{max}\neq \Gamma_0$, treating $m$ as a continuous parameter, we obtain the optimal excitation number $N-m_*$ where
\begin{equation}\label{eq:m_optimal}
    m_* = \frac{N}{2}\left(1+\frac{2}{N}+\frac{N-1}{N}\frac{\Gamma_0}{\Gamma_0-\Gamma_\text{max}}\right).
\end{equation}
In the limit of all-to-all interactions, the system reduces to the Dicke model where the largest decay is $\Gamma_\text{max}=N\Gamma_0$. From Eq.~(\ref{eq:m_optimal}), one then recovers the well known result $m_*=N/2$~\citesupp{dicke1954coherencesupp,Gross1982supp,Andreev1980supp}. To obtain a physical state, we can round $m_\star$ to the nearest integer, which does not affect its asymptotic behavior.}

\new{As an example, let us now consider the case of atomic arrays in the limit of large system size $N$. In this limit, the eigenstates of $\boldsymbol{\Gamma}$ are localized in $k$-space (see Sec.~\ref{sec:Delocalized_Decay}) and can thus be approximated by Fourier modes as discussed in Ref.~\citesupp{Clemens2003collectivesupp}. For 2D and 3D arrays in free space, the largest decay rate scales as $\Gamma_\text{max}\sim N^\alpha$ with $\alpha=1/4$ and $\alpha=1/3$ respectively (see Sec.~\ref{app:scaling_Gamma_max}). Hence, in the asymptotic limit of large $N$, the optimal number of excitations is given by
\begin{equation}\label{eq:n_opt_2D_3D}
    m_* = \frac{N}{2}\left[1 - O(N^{-\alpha})\right]
\end{equation}
for any value of $d/\lambda_0$. For one-dimensional arrays, $m_*$ depends on the interatomic distance $d/\lambda_0$. In the asymptotic limit of large $N$, the largest transition rate is well approximated by $\Gamma_\text{max} = \beta\Gamma_0$. Substituting this expression into Eq.~(\ref{eq:m_optimal}) we obtain
\begin{equation}
    m^{\text{1D}}_* = \frac{\beta-2}{2\beta-2}N + \frac{2\beta-1}{2\beta-2},
\end{equation}
which is valid provided $\beta\neq 1$. Through $\beta$, the optimal excitation number for a 1D array depends on the lattice constant and polarization. At small but finite lattice constant, $m^{\text{1D}}_*\approx N/2$ as $\beta\gg1$ [see for instance Fig.~\protect\ref{fig:Numerical_Scaling}(a) in the region where $10^{-3}\lesssim d/\lambda_0\lesssim 10^{-1}$], irrespective of the polarization. At larger spacings, $m^{\text{1D}}_*$ decreases until it reaches zero at $d\gg \lambda_0$ (as $\beta \rightarrow 1$ the maximal decay happens when the system is fully inverted, as expected for non-interacting atoms). This contrasts with the universal behavior shown in Eq.~(\ref{eq:n_opt_2D_3D}) for 2D and 3D arrays. The reason behind this difference is connected to the existence of a critical lattice constant $d_*$ beyond which the collective decay of a 1D array differs significantly from that of higher-dimensional ones~\citesupp{sierra2022dickesupp}.}
\\

\subsection{No scaling law in the absence of delocalized decay}\label{app:reductio}
The scaling law $R_\star \sim N \Gamma_{\text{max}}$ derived for systems in the delocalized regime is not true for arbitrary systems. We now prove that no general scaling law exists that depend solely on $N$ and the spectrum of $\mathbf{\Gamma}$. This can be seen by \textit{reductio ad absurdum}. Let us suppose $R_\star \sim f(N) g(\Gamma_1,\ldots,\Gamma_N)$, where $f$ and $g$ are arbitrary functions. By applying a \new{unitar} transformation on $\mathbf{\Gamma}$, one can obtain a diagonal decoherence matrix $\mathbf{\Gamma^\prime} = \text{diag}(\Gamma_1,\ldots,\Gamma_N)$. Since $\mathbf{\Gamma^\prime}$ has the same spectrum as $\mathbf{\Gamma}$, the new maximal decay rate $R_\star^\prime \sim R_\star$ by assumption. However, $\mathbf{\Gamma^\prime}$ physically describes a system of $N$ independent qubits with decay rates $\Gamma_1,\ldots,\Gamma_N$, respectively. Hence, $R_\star^\prime = \sum_i \Gamma_i = N\Gamma_0$ which is not $\sim R_\star$ in general. Thus, by contradiction, a general scaling law for $R_\star$ of the form $f(N)g(\Gamma_1,\ldots,\Gamma_N)$ does not exist, without assumptions on the system. This does not invalidate our scaling law for translationally invariant systems, as orthogonal transformations of $\mathbf{\Gamma}$ generally break translation symmetry.

\subsection{Decay rate of typical quantum states}
\label{app:typicalstates}
Here, we prove that the decay rate of typical quantum states that are drawn uniformly from the many-body Hilbert space, i.e., via the Haar measure on the unitary group $\mathit{U}(2^N)$, scales linearly with the system size $N$, implying that typical states do not experience collectively-enhanced decay.

We assume that the Hamiltonian in Eq.~\eqref{eq:lindbladME} contains only geometrically local interactions acting on a constant number of qubits, with non-extensive parameter values. As shown in Section~\ref{sec:disorderandmore}, this only shifts the decay rate $R$ by $O(N)$. Thus, in what follows, we omit the Hamiltonian contribution. The average decay rate (over the Haar measure) is
\begin{equation}
    R_{\text{typ}} \equiv \mathbb{E}_{\ket{\psi} \in \text{Haar}}[\braket{\psi|\hat{H}_\Gamma|\psi}] = \frac{1}{2^N}\text{Tr}\hat{H}_\Gamma = \frac{1}{2^N} \sum_{i,j=1}^{N} \Gamma_{ji} \text{Tr}(\sigma_i^+ \sigma_j^-) = \frac{1}{2^N} \sum_{i=1}^{N} \Gamma_{ii} \text{Tr}(\sigma_i^+ \sigma_i^-)= \frac{N\Gamma_0}{2},
\end{equation}
using \new{$\sum_{i=1}^{N}\Gamma_{ii} = N\Gamma_0$} and $\text{Tr}(\sigma_i^+ \sigma_i^-) =  2^{N-1}$ (from the identity acting on the remaining $N-1$ qubits). The value of the typical rate is rather intuitive since the fully excited state $\ket{e}^{\otimes N}$ has a decay rate of $N\Gamma_0$, while the ground state $\ket{g}^{\otimes N}$ has a decay rate of $0$. Next, we show that for any typical (Haar random) state, the decay rate is close to $R_{\text{typ}} = N\Gamma_0/2$, with a fluctuation that vanishes rapidly with $N$. This arises from the concentration of measure~\citesupp{ledoux2001concentrationsupp}. More precisely, we invoke Levy's lemma, which in our context states that for any observable $\hat{O}$ and any $\epsilon \geq 0$~\citesupp{mele2024introductionsupp},
\begin{equation}
    \text{Pr}\left(\left|\braket{\psi|\hat{O}|\psi} - \text{Tr}\hat{O}/2^N \right| \geq \epsilon \right) \leq 2 \exp\left(-\frac{2^N \epsilon^2}{18\pi^3 \Vert \hat{O} \Vert^2} \right),
\end{equation}
where $\ket{\psi}$ is a Haar random state, $\Vert \hat{O} \Vert$ is the spectral norm, and Pr stands for probability. In our case, $\hat{O} = \hat{H}_\Gamma$, so we have
\begin{equation}
    \text{Pr}\left(\left|\braket{\psi|\hat{H}_\Gamma|\psi} - N\Gamma_0/2 \right| \geq \epsilon \right) \leq 2 \exp\left(-\frac{2^N \epsilon^2}{18\pi^3 R_\star^2} \right).
\label{eq:levy_rate}
\end{equation}
Since $R_\star$ is at most $\sim N^2$, the probability of the rate deviating from $R_\text{typ}$ is doubly exponentially suppressed in $N$. Thus, the decay rate of a typical state is $N\Gamma_0/2$, up to a correction linear in $N$ from the Hamiltonian. 

We remark that our conclusion does not only hold for Haar random states, but also more generally for pseudorandom state ensembles known as \textit{$k$-designs}~\citesupp{ambainis2007quantumsupp}, which are statistically indistinguishable from the Haar ensemble up to the first $k$ moments. Such states can emerge naturally from the infinite-temperature dynamics of chaotic Hamiltonians, and are also useful for quantum information applications. Perhaps the most well-known example is the set of $N$-qubit stabilizer (Clifford) states, which form a $3$-design~\citesupp{webb2016cliffordsupp,zhu2017multiqubitsupp}. However, for $k$-designs, the concentration result of Eq.~\eqref{eq:levy_rate} does not hold generally. Using large deviation bounds for $k$-designs~\citesupp{low2009largesupp}, one can show that the probability of the decay rate deviating from $R_{\text{typ}}$ is exponentially suppressed in $N$ (instead of doubly exponentially suppressed).

\section{Atomic arrays in free space}

\new{In this section, we provide additional results on the scaling laws for atomic arrays in free space. This appendix is structured as follows. In Sec.~\ref{app:scaling_Gamma_max}, we prove the analytical scaling of $\Gamma_\text{max}$. We derive the scaling of $\Gamma_\text{max}$ for an arbitrary $\delta$-dimensional lattice in a $D\geq\delta$ dimensional free space in Sec.~\ref{app:Higher_Dimensions}. In Sec.~\ref{sec:Delocalized_Decay}, we show that the decay is delocalized for finite but large arrays. In Sec.~\ref{app:Disorder} we show that the scaling laws are robust against position disorder in experimental realizations. Finally, we estimate the typical timescale of superradiant emission in Sec.~\ref{app:Burst_Time}. In this section, we consider $\Gamma_{ii}=\Gamma_0$ for the sake of clarity. Nevertheless, the results are valid in general as changes to the diagonal elements of $\boldsymbol{\Gamma}$ do not change $R_\star$, as demonstrated in Sec.~\ref{sec:disorderandmore}.}

\new{\subsection{Proof of the scaling of the largest transition rate for atomic arrays in free space}\label{app:scaling_Gamma_max}}

We consider ensembles of $N$ two-level atoms arranged in 1D, 2D and 3D ordered arrays \new{in free space} with lattice constant $d$. Tracing out the vacuum electromagnetic modes within the Born-Markov approximation, one obtains an effective master equation for the atomic dynamics of the same form as Eq.~(\ref{eq:lindbladME})\new{~\citesupp{Lehmberg1970asupp}, i.e.,
\begin{equation}
    \dot{\hat{\rho}} = -\frac{i}{\hbar}[\hat{H},\hat{\rho}] + \sum_{i,j=1}^{N} \Gamma_{ij} \left(\sm_i \hat{\rho} \sp_j - \frac{1}{2}\{\sp_j \sm_i, \hat{\rho} \} \right),\quad\text{where } \hat{H}=\hbar\omega_0\sum_i^N\sz_i+\hbar\sum_{i\neq j}^{N}J_{ij}\sp_i \sm_j.
\label{eq:atomicME}
\end{equation}
The coherent} and dissipative couplings between atoms $i$ and $j$ are \new{correspondingly} given by \new{the real and imaginary parts} of the field propagator between them (projected in the direction of the atomic transition dipole element), $\textbf{G}(\textbf{r}_i,\textbf{r}_j,\omega_0)\equiv \textbf{G}(\textbf{r}_{ij},\omega_0)$, where $\textbf{r}_{ij}=\textbf{r}_i-\textbf{r}_j$ is the relative position between the atoms. \new{The coupling rates read~\citesupp{AsenjoGarcia2017PRXsupp}
\begin{equation}
\begin{split}
    J_{ij}&= -\frac{3\pi \Gamma_0}{k_0} \hat{\dpp}^{*}\cdot \text{Re}\,\textbf{G}(\textbf{r}_i,\textbf{r}_j,\omega_0) \cdot \hat{\dpp},\\
    \Gamma_{ij}&= \frac{6\pi \Gamma_0}{k_0} \hat{\dpp}^{*}\cdot \text{Im}\,\textbf{G}(\textbf{r}_i,\textbf{r}_j,\omega_0) \cdot \hat{\dpp}, \label{eq:Gamma}
\end{split}
\end{equation}
where $\w_0=ck_0$ is the atomic resonance frequency, $\Gamma_0=\w_0^3|\dpp|^2/3\pi\hbar\epsilon_0 c^3$ is the single-atom spontaneous emission rate, and $\dpp$ is the dipole matrix element of the atomic transition.} In 3D vacuum, the electromagnetic Green's tensor reads~\citesupp{jacksonsupp}
\begin{equation}\label{eq:green}
\begin{split}
    \textbf{G} (\textbf{r},\w_0) \equiv& \frac{e^{\ii k_0 r}}{4\pi k_0^2r^3}\Big[(k_0^2r^2+\ii k_0 r-1)\mathds{1}+(-k_0^2r^2-3\ii k_0r+3)\frac{\textbf{r}\otimes\textbf{r}}{r^2}\Big],\,\,\, \text{where}\,\,r=|\textbf{r}|.
\end{split}
\end{equation}

\new{To find analytical expressions for} the scaling of \new{$\Gamma_\text{max}$, it is convenient to study an infinite array. In the limit of large system size, the jump operators can be well approximated by Fourier modes~\citesupp{Clemens2003collectivesupp} so that Eqs. \eqref{eq:Gamma} are diagonal in $k$-space as confirmed numerically in several works~\citesupp{Bettles2016supp,AsenjoGarcia2017PRXsupp}. In the following, we thus assume that the single-excitations eigenmodes of a large array can be described by spin waves with well defined quasimomentum $\kk$~\citesupp{AsenjoGarcia2017PRXsupp}. In momentum space, the coherent and dissipative} transition rates read
\begin{equation}
\begin{split}
    \new{J(\kk)}&\new{= -\frac{3\pi\Gamma_0}{k_0} \hat{\dpp}^{*}\cdot \text{Re}\,\tilde{\textbf{G}}(\kk) \cdot \hat{\dpp},} \\
    \Gamma(\kk)&= \frac{6\pi\Gamma_0}{k_0} \hat{\dpp}^{*}\cdot \text{Im}\,\tilde{\textbf{G}}(\kk) \cdot \hat{\dpp}, \label{eq:gf_bz}
\end{split}
\end{equation}
where $\tilde{\textbf{G}}(\kk)=\sum_j e^{-i\kk\cdot\rb_j}\textbf{G}(\rb_j,\omega_0)$ is the \new{discrete} Fourier transform of the Green's function in Eq.~\eqref{eq:green}. 

\new{Below, we examine each array dimensionality separately. Since $J(\kk)$ does not contribute to $R_\star$, we restrict the analysis to $\Gamma(\kk)$ and focus our attention on the appearance of divergences that will drive the behavior of $\Gamma_\text{max}$. We simplify the analysis by focusing on the dominant dependence of $\Gamma_\text{max}$ with system size and expressing it as $\Gamma_\text{max}=\beta N^\alpha\Gamma_0$, where $\beta$ has no explicit dependence on $N$. In the extreme cases of infinitely small or infinitely large lattice constants, the scaling parameters also depend on the order of the limits $N\rightarrow\infty$ and $d\rightarrow 0$ (or $d\rightarrow \infty$), as discussed later.}

% \subsubsection{One-dimensional ordered arrays}
\new{{\bf One-dimensional ordered arrays.} We consider an infinite 1D array along the $z$ direction. }The transition rate of a spin-wave with momentum $k_z$ \new{is uniquely defined inside the first Brillouin zone} and can be computed from Eq.~\eqref{eq:gf_bz}. \new{An analytical expression can be obtained making use of a Weyl expansion after expressing Eq.~\eqref{eq:green} in terms of spherical waves, and then transforming the expression to reciprocal space via the Dirac delta representation to find (see Ref.~\citesupp{AsenjoGarcia2017PRXsupp} for the details)}
\begin{equation}\label{eq:Gamma_1D_k}
\begin{split}
    \frac{\Gamma_\text{1D}^\parallel(k_z)}{\Gamma_0} &= \frac{3\pi}{2k_0d}\sum_{g_z} \left(1-\frac{(k_z+g_z)^2}{k_0^2}\right),\\
    \frac{\Gamma_\text{1D}^\perp(k_z)}{\Gamma_0} &= \frac{3\pi}{4k_0d}\sum_{g_z} \left(1+\frac{(k_z+g_z)^2}{k_0^2}\right),
\end{split}
\end{equation}
where the summation runs over reciprocal lattice vectors $g_z = 2\pi n/d$ for $n\in \mathbb{Z}$ that satisfy the condition $|g_z+k_z|\leq k_0$. Here, the superscripts $\{^\parallel,^\perp\}$ refer to atoms with polarization parallel and perpendicular to the array axis, respectively. 

\new{The decay rates are always finite. The lack of any divergence as a function of $k_z$ implies that the maximum transition rates do not scale with $N$. To make this connection explicit, we approximate the transition rates of a finite 1D array of $N$ atoms by sampling Eq.~\eqref{eq:Gamma_1D_k} with an equally spaced grid such that $\Gamma_\mu=\Gamma_\text{1D}(k_\mu)$, with $k_\mu=-\pi/d + 2\pi\mu/(N+1)$ for $\mu\in \{1,\dots,N\}$. For simplicity, let us assume that $0<d<\lambda_0/2$ so that the sum over reciprocal lattice vectors disappears. The maximum transition rate is therefore the closest point to the absolute maximum of Eq.~\eqref{eq:Gamma_1D_k}, located at $k_z=0$ (for parallel polarization) and $k_z=k_0$ (for perpendicular polarization). Then, the maximum transition rates are
\begin{equation}
\begin{split}
        \frac{\Gamma_\text{max}^\parallel}{\Gamma_0} &= \frac{3\pi}{2k_0d}\left(1-\varepsilon^2\right),\\
        \frac{\Gamma_\text{max}^\perp}{\Gamma_0} &=\frac{3\pi}{4k_0d}\left(1+(1-\varepsilon)^2\right),
\end{split} \label{eq:expansion_1D}
\end{equation}
where $\varepsilon$ is the distance between the absolute maximum and the closest grid point. For parallel polarization $\varepsilon=\{0,\pi/k_0d(N+1)\}$ depending on the parity of $N$, and for perpendicular polarization,  $\varepsilon k_0d(N+1)=2\pi\,\text{frac}(k_0d(N+1)/2\pi)$, with $\text{frac}(x)=x - \lfloor x \rfloor$ being the fractional part of $x>0$. This means that the constant term is the leading term as a function of $N$, so $\alpha^\text{(1D)}=0$ and $\beta^\text{(1D)}= 3\pi/2k_0d$. For a finite $d\geq\lambda_0/2$, the number of terms in the sum increases with $d$ and so does $\beta^\text{(1D)}d$, keeping the dependence on $N$ unchanged.}

\new{The limiting cases of zero and infinite lattice constants have to be treated separately. When taking the limit $d \to\infty$ in Eq.~\eqref{eq:Gamma_1D_k}, the sum} over reciprocal lattice vectors extends up to \new{$g_\text{max}=2\pi n_*^\pm/d$, where $n_*^\pm=\pm\lfloor d/\lambda_0\mp k_zd/2\pi\rfloor$. Then,
\begin{equation}
\begin{split}
    \frac{\Gamma_\text{1D}^\perp(k_z)}{\Gamma_0} &= \lim_{d\to\infty} \frac{3\pi}{4k_0d}\sum_{g_z} \left(1+\frac{(k_z+g_z)^2}{k_0^2}\right) = \lim_{d\to\infty} \frac{3\pi}{4k_0d} \sum_{m=n_*^-}^{m=n_*^+} \left(1+\frac{(k_z+2\pi m/d)^2}{k_0^2}\right) \\
    &= \frac{3\pi}{4k_0d}\int_{-d/\lambda_0-k_zd/2\pi}^{d/\lambda_0-k_zd/2\pi} \left(1 +  \left(k_zd/2\pi + x\right)^2 \lambda_0^2/d^2\right) dx = 1.
\end{split}
\end{equation}
A similar argument follows for parallel polarization, so that} one recovers the non-interacting solution $\Gamma_\text{1D}^{\parallel}(k_z)=\Gamma_\text{1D}^{\perp}(k_z)=\Gamma_0,\,\forall k_z$, \new{and therefore $\alpha^\text{(1D)}=0$ and $\beta^\text{(1D)}= 1$ when $d \to\infty$. Note that we have first taken the limit of $N\to \infty$ and then that of $d \to\infty$}. Since Eq.~\eqref{eq:Gamma_1D_k} does not diverge in $k$-space, inverting the order of the two limits leaves the result unchanged. 

We now consider the Dicke limit of an infinitely small lattice constant. We start from Eq.~(\ref{eq:gf_bz}) and, separating the diagonal term, we write (for perpendicular polarization)

\begin{equation}\label{eq:Gamma_1D_Dicke}
\begin{split}
    \frac{\Gamma_\text{1D}^\perp(k_z)}{\Gamma_0} &= 1 + \lim_{d\to 0} \frac{3}{4(k_0d)^3}\text{Im}\left\{\sum_{i\neq j=1}^N \frac{e^{\ii k_0 d |i-j|}}{|i-j|^3}\frac{e^{\ii k_z d (i-j)}}{N}\left(1-\ii k_0 d |i-j|-k_0^2d^2 |i-j|^ 2\right)\right\} \\ &\new{=1 + \frac{1}{N}\!\! \sum_{i\neq j=1}^{N}\cos(k_zd |i-j|)= \frac{1}{N}\!\! \sum_{i, j=1}^{N}\cos(k_zd |i-j|)=N\delta (k_z)}, 
\end{split}
\end{equation}
\new{where we have expanded around $k_0d =0$ and used that the} contribution proportional to $\sin[k_z d(i-j)]$ vanishes \new{due to symmetry. We then have $\Gamma_\text{max}=N\Gamma_0$, so that $\alpha^\text{(1D)}=1$ and $\beta^\text{(1D)}= 1$ when $d \to 0$}, recovering Dicke's scaling. A similar argument follows for parallel polarization, yielding the same result. For vanishing lattice constants, we cannot exchange the order of the limits as Eq.~\eqref{eq:Gamma_1D_Dicke} diverges for $N\rightarrow \infty$. This can be seen from Eq.~\eqref{eq:expansion_1D}, as the constant term is no longer the dominant one when the small lattice constant limit is taken before $N\to\infty$, as then $\varepsilon\to\infty$. The transition between the two limiting cases happens then when $\varepsilon \sim 1$, i.e. for $N\sim (k_0d)^{-1}$ or $L=Nd\sim\lambda_0$.

\begin{figure}
    \centering
    \includegraphics[width=\textwidth]{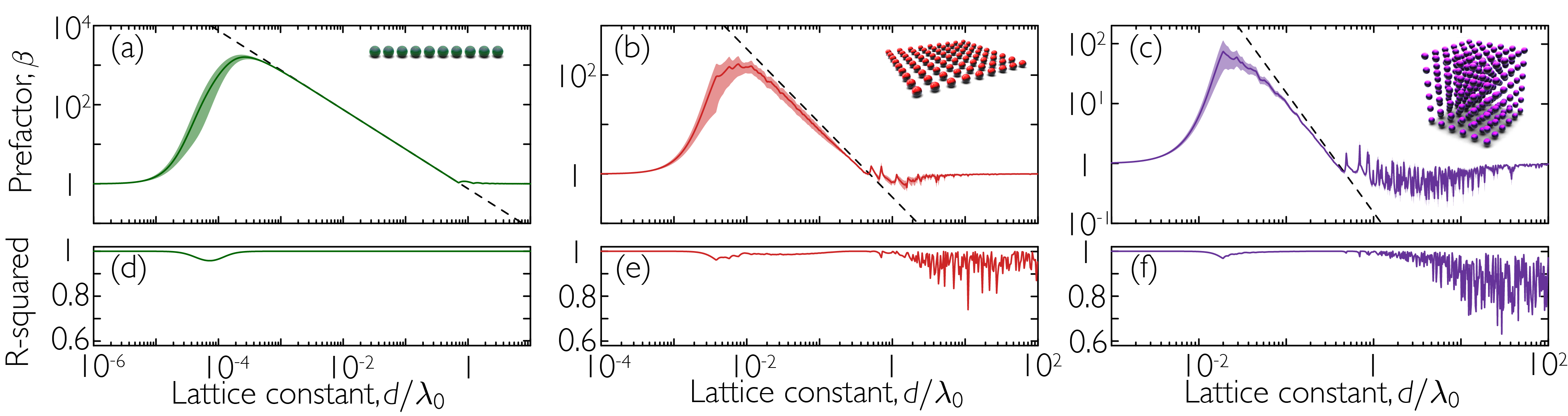}
    \caption{\new{Scaling prefactor (a)-(c) and R-squared value (d)-(f) of the largest transition rate with system size (obtained from a best fit to $\Gamma_\text{max}=\beta N^\alpha \Gamma_0$) as a function of lattice constant for 1D (left panels), 2D (central panels) and 3D (right panels) atomic arrays. The atoms form a square lattice and are polarized parallel to one axis of the array. The fits are done over a region $N_\text{1D}\in [2, N^\text{max}_\text{1D}]$, where $N^\text{max}_\text{1D}=\{30000, 250, 40\}$ for 1D, 2D, and 3D, respectively, by sampling seven equally spaced points. The shaded colored area around the lines represents the 1$\sigma$ confidence interval. The dashed lines show the scaling with lattice constant of the prefactor $\beta$ for large arrays. These are $\beta^{(\text{1D})}=3\pi/2k_0d$, $\beta^{(\text{2D})}=3\sqrt{\pi}/2(k_0d)^{3/2}$, and $\beta^{(\text{3D})}=3/5(k_0d)^2$, as discussed in Section~\ref{app:scaling_Gamma_max}.}}
    \protect\label{fig:Numerical_Scaling}
\end{figure}

These analytical limits are recovered numerically by fitting $\Gamma_\text{max}=\beta N^\alpha\Gamma_0$, as shown by \new{Figs.~\ref{fig3} and~\protect\ref{fig:Numerical_Scaling}}. In particular, both $\alpha$ and $\beta$ converge to the expected Dicke and non-interacting limits at small and large interatomic separation\new{, respectively}. In the intermediate regime of \new{sub-wavelength separation $1/N\lesssim d/\lambda_0\lesssim 1/2$}, the parameters $\beta$ and $\alpha$ scale as for the asymptotic case in Eq.~\eqref{eq:Gamma_1D_k}.

\new{{\bf Two-dimensional ordered arrays.} We now consider a 2D array extended in the $xy$ plane. The transition rate} of a spin-wave with momentum $\kk=(k_x,k_y)$ in the 2D plane \new{can be written as~\citesupp{AsenjoGarcia2017PRXsupp}}
\begin{subequations}
\begin{eqnarray}
    \frac{\Gamma_\text{2D}^\parallel (\kk)}{\Gamma_0} &=& \frac{3\pi}{k_0^3d^2}\sum_{\mathbf{g}} \frac{k_0^2-|(\kk+\mathbf{g})\cdot \hat{\dpp}|^2}{\sqrt{k_0^2-|\kk+\mathbf{g}|^2}},\label{eq:Gamma_2D_k_parallel}\\
    \frac{\Gamma_\text{2D}^\perp (\kk)}{\Gamma_0} &=& \frac{3\pi}{k_0^3d^2}\sum_{\mathbf{g}} \frac{|\kk+\mathbf{g}|^2}{\sqrt{k_0^2-|\kk+\mathbf{g}|^2}},\label{eq:Gamma_2D_k_perp}
\end{eqnarray}
\end{subequations}
where the sum extends over all reciprocal lattice vectors, $\mathbf{g}=2\pi(n_x,n_y)^T/d$ for $n_x,n_y\in\mathbb{Z}$ that satisfy the condition $|\kk+\mathbf{g}|<k_0$. Both equations diverge for $|\kk+\mathbf{g}|\rightarrow k_0$, but Eq.~(\ref{eq:Gamma_2D_k_parallel}) avoids the divergence along the polarization direction $\hat{\dpp}$. 

\begin{figure}
    \centering
    \includegraphics[width=0.65\columnwidth]{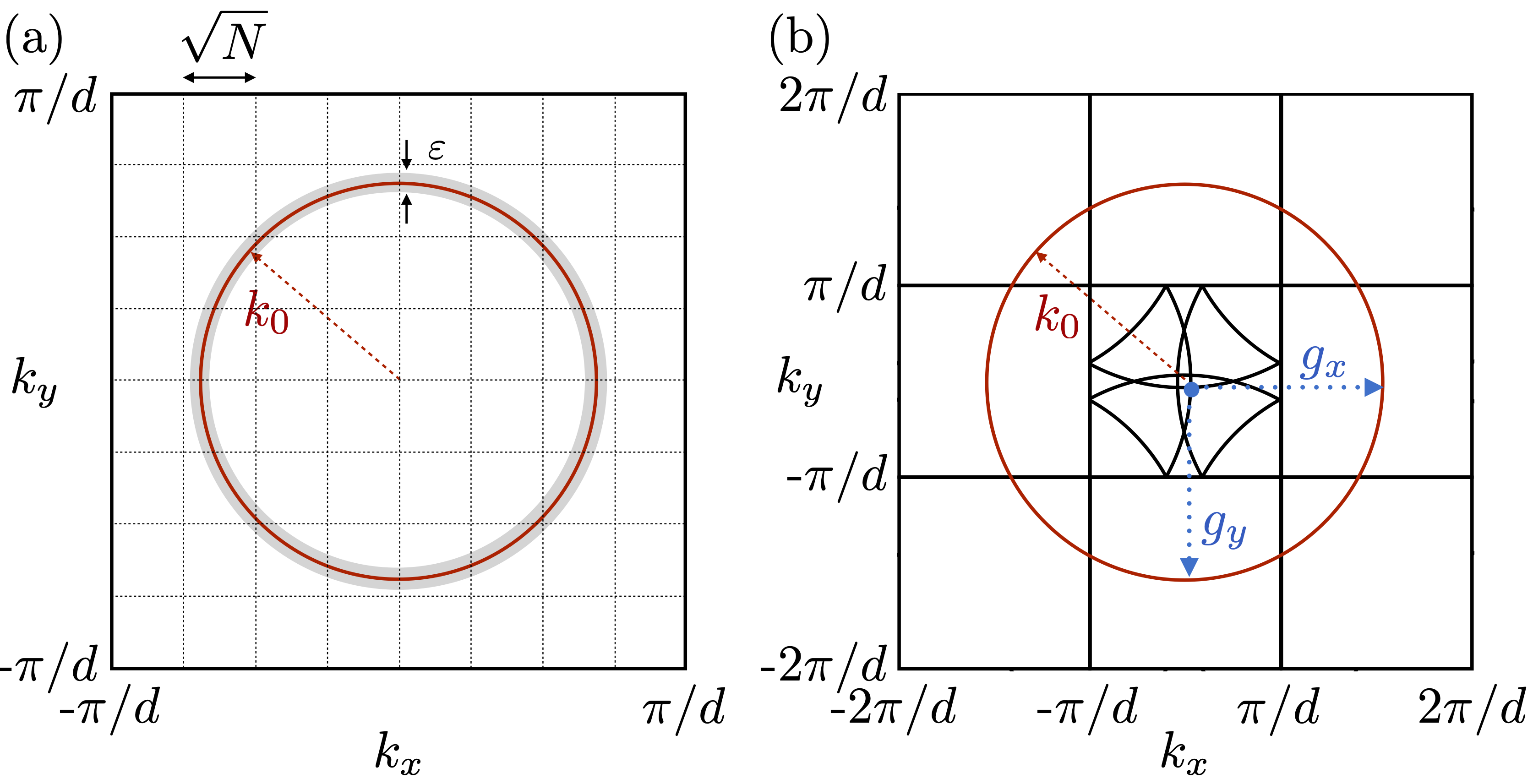}
    \caption{\new{Description of the divergences in $k$-space for a 2D array.} (a) For $d<\lambda_0/2$ the light cone (red circle) is contained in the first Brillouin zone, $k_0<\pi/d$. The intersection of the $\sqrt{N}\times\sqrt{N}$ black dashed grid lines are the allowed values of  $\kk$ for a finite $N$ atom array. The gray region around the light cone is excluded to remove the divergence. (b) For $d>\lambda_0/2$, the light cone (red circle) is larger than the first Brillouin zone. The black intersecting lines at the center are obtained by folding the light line into the first Brillouin zone. The point $\kk$ where two lines intersect (blue circle) can be reached by two distinct scattering processes represented by the vectors $\mathbf{g}_{x,y}$ and satisfying the condition $|\kk+\mathbf{g}_\alpha|=k_0$ for $\alpha=\{x,y\}$ (blue dotted arrows).}
    \protect\label{fig:Brillouin_Zone}
\end{figure}

To understand how this divergence translates into the asymptotic scaling $\Gamma_\text{max}/\Gamma_0 \propto N^{1/4}$, \new{let us} first consider the case of \new{$0<d<\lambda_0/2$, where $k_0<\pi/d$} and the only term that contributes to the sums in Eqs.~(\ref{eq:Gamma_2D_k_parallel}-\ref{eq:Gamma_2D_k_perp}) is $\mathbf{g}=0$. For an array of $N$ atoms, \new{we can approximate the transition rates by sampling} $\Gamma(\kk)$ on a finite $\sqrt{N}\times\sqrt{N}$ grid in momentum space. \new{As shown in  Fig.~\protect\ref{fig:Brillouin_Zone}(a), the wavevectors with larger transition rates are those closer} to the divergence \new{such that $|\textbf{k}|=k_0(1-\varepsilon)$, where $\varepsilon\leq 2\pi/k_0d(\sqrt{N}+1)$ (for parallel polarization the direction is fixed to be close to perpendicular to $\hat{\dpp}$).}  Plugging this wavevector into the above expressions, we \new{readily find}
\new{\begin{equation}
    \begin{split}
        \frac{\Gamma_\text{max}^\parallel (\kk)}{\Gamma_0} &=\frac{3\pi}{k_0^2d^2} \frac{1}{\sqrt{1-(1-\varepsilon)^2}} \simeq \frac{3\pi}{\sqrt{2}k_0^2d^2}\varepsilon^{-1/2} + O(\varepsilon^{1/2})\simeq \frac{3\sqrt{\pi}}{2(k_0d)^{3/2}}N^{1/4},\\
        \frac{\Gamma_\text{max}^\perp (\kk)}{\Gamma_0} &=\frac{3\pi}{k_0^2d^2} \frac{(1-\varepsilon)^2}{\sqrt{1-(1-\varepsilon)^2}} \simeq \frac{3\pi}{\sqrt{2}k_0^2d^2}\varepsilon^{-1/2} + O(\varepsilon^{1/2})\simeq \frac{3\sqrt{\pi}}{2(k_0d)^{3/2}}N^{1/4}, 
    \end{split}\label{eq:expansion_2D}
\end{equation}}
\new{so that} $\alpha^\text{(2D)}=1/4$ and \new{$\beta^\text{(2D)}= 3\sqrt{\pi}/2(k_0d)^{3/2}$}. For $d_0>\lambda_0/2$, $k_0>\pi/d$, and higher-order scattering processes (with $\mathbf{g}\neq0$) are allowed. For some values $\kk$ within the first Brillouin zone there can be more than one value of $\mathbf{g}$ such that $|\kk+\mathbf{g}|=k_0$, as illustrated in Fig.~\protect\ref{fig:Brillouin_Zone}(b). These special values of $\kk$ are those points where the light line intersects with itself once folded into the first Brillouin zone [black curves in Fig.~\protect\ref{fig:Brillouin_Zone}(b)]. The number of distinct solutions for $\mathbf{g}$ to the equation $|\kk+\mathbf{g}|=k_0$ agrees with the number of intersecting lines at $\kk$. This multiplicity of solutions has an effect on the values of $\Gamma_\text{max}$ for a finite array. In particular, as $d/\lambda_0$ becomes larger, $\Gamma_\text{max}$ suddenly grows any time the maximum number of lines intersecting in one point in the first Brillouin zone increases. This leads to the peaks observed in the region $d>\lambda_0/2$ in Fig.~\ref{fig3}. \new{These features are generic except for 1D arrays with parallel polarization, as light emission in the direction of the chain is forbidden. Nevertheless,} the scaling holds regardless of the terms in the sum (in the $N\rightarrow\infty$ limit).

\new{We now proceed to analyze the two limit cases of zero and infinite inter-particle separation. Because of the pole in reciprocal space, the order of the limits $N\rightarrow\infty$ and $d/\lambda_0\rightarrow\infty$ matters. If one takes the limit of large separation in Eqs.~(\ref{eq:Gamma_2D_k_parallel}-\ref{eq:Gamma_2D_k_perp}) (where the $N\rightarrow\infty$ has already been taken), the scaling of $\Gamma_\text{max}$ does not change as $\varepsilon$ is strictly decreasing both as a function of $N$ and $d$, so the leading term of the expansion is the same. On the contrary, by taking the large separation limit first at the level of Eq.~\eqref{eq:Gamma}, one recovers the limit of independent atoms $\Gamma_\text{2D}^\parallel (\kk)=\Gamma_\text{2D}^\perp (\kk)=\Gamma_0$ regardless of $N$, so that $\alpha^\text{(2D)}=0$ and $\beta^\text{(2D)}= 1$. The transition between these two limits can be seen directly from Eq.~\eqref{eq:expansion_2D}, as the first term is the only one for which the limit is not independent of the order, changing behavior at $N^\text{crit}= 16(k_0d)^6/81\pi^2$ (when $\Gamma_\text{max}>\Gamma_0$ asymptotically). The limit of infinitely small separation is similar to the 1D case. By taking first the limit $d\to 0$ in Eq.~\eqref{eq:gf_bz} one finds that $\Gamma_\text{2D}^\parallel (\kk)=\Gamma_\text{2D}^\perp (\kk)=N\Gamma_0\delta(\kk)$. We then have that $\Gamma_\text{max}=N\Gamma_0$, so that $\alpha^\text{(2D)}=1$ and $\beta^\text{(2D)}= 1$ when $d \to 0$, recovering Dicke's scaling. On the contrary, Eq.~\eqref{eq:expansion_2D} shows that taking the limit $N\to\infty$ first does not change the normal scaling as long as $\varepsilon\to 0$. The transition between the two limiting cases happens then when $\varepsilon \sim 1$, i.e. when $N\sim (k_0d)^{-2}$ or $L=N_\text{1D}d\sim\lambda_0$.}

\new{{\bf Three-dimensional ordered arrays.} 3D arrays are special, as infinitely-large arrays have a zero transition rate at any point of the Brillouin zone, except for the light line. Therefore, the transition rates become a Dirac delta, showing a divergence at $\kk=\kk_0$.} Here, we discuss how this divergence is approached as $N$ increases. The maximum transition rate is numerically shown to scale as $\Gamma_\text{max}/\Gamma_0\sim N^{1/3}$. To prove this result, one can introduce a regularization factor  $\Delta\rightarrow 0^+$ that controls the divergence of $\Gamma_\text{3D}(\kk)$, and take the limit $N\rightarrow\infty$~\citesupp{Antezza2009supp,Brechtelsbauer2021supp}. One then obtains~\citesupp{sierra2022dickesupp}
\begin{equation}\label{eq:Gamma_3D_k}
    \frac{\Gamma_\text{3D}(\kk)}{\Gamma_0} = \frac{6\pi}{k_0d^3}\sum_{\mathbf{g}} \frac{\Delta (k_0^2-|(\kk+\mathbf{g})\cdot\hat{\dpp}|^2)}{(k_0^2-|(\kk+\mathbf{g})|^2)^2+\Delta^2k_0^4},
\end{equation}
where $\mathbf{g}=2\pi(n_x,n_y,n_z)/d$ for $n_x,n_y,n_z\in\mathbb{Z}$ and the sum is extended to all values of $\mathbf{g}$ that satisfy the condition $|\kk+\mathbf{g}|<k_0$. 

\new{We derive the asymptotic scaling using an argument analogous to that employed for 2D arrays. For simplicity, we consider the regime $0<d<\lambda_0/2$, where the sum over reciprocal lattice vectors vanishes. We sample the Brillouin zone on a $\sqrt[3]{N}\times\sqrt[3]{N}\times\sqrt[3]{N}$ grid, so that the wavevectors with larger transition rates are those closer to the divergence, with $|\kk|=k_0(1-\varepsilon)$. Assuming $\Delta\sim\varepsilon\leq 2\pi/k_0d(\sqrt[3]{N}+1)$, we find
\begin{equation}
        \frac{\Gamma_\text{max} (\kk)}{\Gamma_0} \sim\frac{6\pi}{k_0^3d^3} \frac{\varepsilon}{(1-(1-\varepsilon)^2)^2+\varepsilon^2} = \frac{6\pi}{5k_0^3d^3}\varepsilon^{-1} + \frac{24\pi}{125k_0^3d^3} + O(\varepsilon)\sim \frac{3}{5k_0^2d^2}N^{1/3},\label{eq:expansion_3D}
\end{equation}
so that} $\alpha^\text{(3D)}=1/3$ and \new{$\beta^\text{(3D)}=3/5(k_0d)^{2}$}. Similar arguments as for 2D also recover both the Dicke and non-interacting limits. \new{The transition between non-interacting and array behavior happens also when the first order of the expansion is of order unity, so that $N^\text{crit}=125(k_0d)^6/27$. The transition between the Dicke and array regions happens when $\varepsilon$ is of order unity, so that $N\sim (k_0d)^{-3}$ or $L=\sqrt[3]{N}d\sim\lambda_0$.}

\new{These arguments can be easily generalized to different lattice geometries. The scalings presented above are universal to any 2D array both in the asymptotic limit and for finite $N$, as shown numerically in Figs.~\ref{fig3} and~\protect\ref{fig:Numerical_Scaling} by fitting $\Gamma_\text{max}=\beta N^\alpha\Gamma_0$. Moreover, this figure shows that we numerically recover the analytical limits predicted in this section. }

\new{\subsection{Scaling of the largest transition rate in arbitrary vacuum dimension}\label{app:Higher_Dimensions}}

\new{In this appendix, we compute the scaling of the largest transition rate $\Gamma_\text{max}$ for a $D$-dimensional regular lattice in $\delta$-dimensional vacuum. As described in the previous appendix, corresponding to $\delta=3$, the scaling of $\Gamma_\text{max}$ is determined by the divergence of the imaginary part of the Green's function in momentum space.  Here, we consider only the scalar Green's function $g(\bold{r})$, as the vectorial part only reduces the divergence for certain polarization directions but never removes it completely, leaving the scaling of $\Gamma_\text{max}$ unchanged. The scalar Green's function in $\delta$-dimensions is the fundamental solution of the wave equation, i.e., 
\begin{equation}\label{eq:Helmholtz_Eq}
    \nabla^2 g(\bold{r}) +k_0^2 g(\bold{r}) = \delta(\bold{r}).
\end{equation}
The atom-atom dissipative coupling is proportional to the imaginary part of the Green's function, namely $\gamma_{ij} \propto \text{Im}\{g(\bold{r}_i-\bold{r}_j)\}$. Since we only consider the scalar part, we have denoted the dissipative interaction rate as $\gamma_{ij}$ (instead of $\Gamma_{ij}$). By taking the Fourier transform of the scalar Green's function defined in the atomic lattice, we find
\begin{equation}\label{eq:Green_f_eq}
    \gamma(\bold{q}) \propto - \text{Im} \left\{ \sum\limits_{\textbf{r}\in\text{lattice}} \int \,d^\delta \textbf{k} \frac{1}{k^2-k_0^2}e^{i(\textbf{k}-\textbf{q})\cdot\textbf{r}}\right\}.\\
\end{equation}
The divergence of Eq.~\eqref{eq:Green_f_eq} in momentum space determines the scaling of the maximum transition rate with the system size as described in Appendix~\ref{app:scaling_Gamma_max}. This divergence depends on the dimensionality $D$ of the lattice and the dimension of physical space $\delta$. We distinguish three different cases: (i) $D<\delta - 1$, (ii) $D=\delta - 1$, and (iii) $D=\delta$.}

\new{(i) For $D<\delta - 1$, we separate the $\delta$-dimensional momentum $\kk$ into a $(\delta-1)$-dimensional component, $\tilde{\kk}$, and a $1$-dimensional component with norm $k_1$. Assuming the lattice to lie in the hyperplane defined by $r_1=0$ and orthogonal to the direction of $k_1$, we write Eq.~(\ref{eq:Green_f_eq}) as
\begin{equation}\label{eq:Green_eq_lattice_diff_by_more_than_1}
    \gamma(\bold{q}) \propto -\text{Im}\! \left\{ \sum_{\textbf{r}\in\text{lattice}}\! \int \!\!\text{d}^{\delta-1} \tilde{\kk}  \int\!\! \text{d}k_1\,\frac{e^{\ii(\tilde{\kk}-\textbf{q})\cdot\textbf{r}}e^{\ii k_1r_1}}{\tilde{k}^2+k_1^2-k_0^2} \right\}.
\end{equation}
The integral over $k_1$ can be computed in the complex plane. It is necessary to distinguish the two cases $\tilde{k}^2<k_0^2$ and $\tilde{k}^2>k_0^2$.  Let us start from the former. Computing the integral over $k_1$, using the Dirac delta representation in $D$-dimensions,
\begin{align}\label{eq:identity1}
\sum\limits_{\textbf{r}\in\text{lattice}}e^{i(\tilde{\kk}-\textbf{q})\cdot\textbf{r}} = \left(\frac{2\pi}{d}\right)^D\sum\limits_{\substack{\textbf{g}} }\delta^{(D)}(\tilde{\kk}-\textbf{q}-\textbf{g}),
\end{align}
where $\textbf{g}$ is a vector of the reciprocal lattice, we obtain
\begin{equation}\label{eq:Green_eq_lattice_diff_by_more_than_1_contd_2}
    \gamma(\bold{q}) \propto \sum\limits_{\textbf{g}} 
    \int\limits_{0}^{\rho_+}\!\!\text{d}\rho \ \frac{\rho^{\delta-D-2}}{\sqrt{k_0^2-|\textbf{q}+\textbf{g}|^2-\rho^2}},
\end{equation}
where we changed variables to spherical coordinate using the equation for the surface area of a $D$-dimensional sphere and $\rho_+ \equiv \sqrt{k_0^2-|\textbf{q}+\textbf{g}|^2}$. The remaining integral in Eq.~(\ref{eq:Green_eq_lattice_diff_by_more_than_1_contd_2}) can be easily computed and we finally obtain
\begin{equation}\label{eq:Green_eq_lattice_diff_by_more_than_1_contd_3}
   \gamma(\bold{q}) \propto \sum\limits_{\textbf{g}}(k_0^2-|\textbf{q}+\textbf{g}|^2)^{\frac{\delta-D-2}{2}}.
\end{equation}
This expression does not diverge. Even though we have derived this result for $k^2<k_0^2$, it also holds for $k^2>k_0^2$. Thus, we conclude that for $D < \delta - 1$, the brightest transition rate does not scale with $N$.}

\new{(ii) For $D = \delta - 1$, we can follow similar steps until Eq.~(\ref{eq:Green_eq_lattice_diff_by_more_than_1_contd_2}), to obtain,
\begin{equation}\label{eq:Green_eq_lattice_diff_by_1}
    \gamma(\bold{q}) \propto  \sum\limits_{\substack{\textbf{g}}} \frac{1}{\sqrt{k_0^2-|\textbf{q}+\textbf{g}|^2}}.
\end{equation}
Hence, we conclude that $\gamma_\text{max} \sim \sqrt{N_{1D}} = N^{\frac{1}{2(\delta-1)}}$.}

\new{(iii) For $D = \delta$, we use Eq.~(\ref{eq:identity1}) and, as for the case of 3D arrays in Appendix~\ref{app:scaling_Gamma_max}, we introduce a regularization factor $i\Delta \to 0^+$ to avoid the divergence at $|\textbf{q}+\textbf{g}| \to k_0$. We then obtain
\begin{equation}\label{eq:Green_eq_lattice_same}
\gamma(\bold{q})\propto\sum\limits_{\substack{\textbf{g}}} \frac{\Delta k_0^2}{(|\textbf{q}+\textbf{g}|^2-k_0^2)^2+(\Delta k_0^2)^2}.
\end{equation}
We can isolate the divergence by considering a momentum close to the light line, such that $|\textbf{q}+\textbf{g}| = k_0(1-\epsilon)$. By taking $\epsilon = C\lambda_0/dN^{1/D}$ (where C is a constant) and imposing that $\Delta\sim\epsilon$ near the divergence in the limit that $\epsilon\to 0^+$, we find that the brightest decay rate scales as $\gamma_\text{max}\sim N_\text{1D} = N^{\frac{1}{\delta}}$.}

\new{\subsection{Delocalized decay in finite atomic arrays}\label{sec:Delocalized_Decay}}

\begin{figure}
    \centering
    \includegraphics[width=\columnwidth]{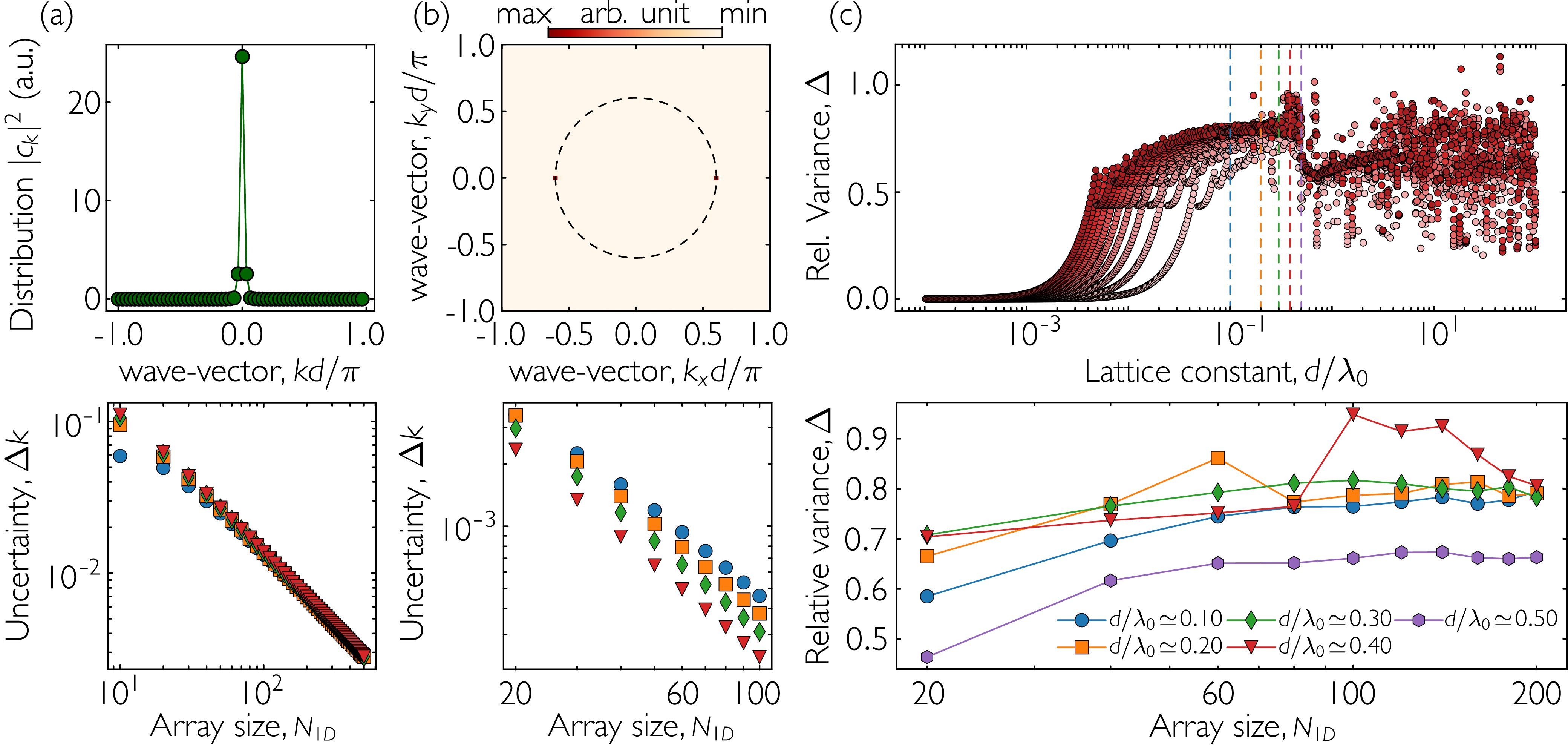}
    \caption{\new{Delocalization of the brightest jump operator for 1D and 2D arrays (polarized along the array axis). (a) 1D arrays: (top) $k$-space distribution $|\alpha^{(1)}_k|^2$ of the brightest jump operators for 60 atoms with $d=0.3\lambda_0$ (principal eigenvector of $\boldsymbol{\Gamma}$) and (bottom) its variance $\Delta k$ as a function of the array size $N_\text{1D}$. (b) 2D arrays: (top)   $k$-space distribution $|\alpha^{(1)}_\kk|^2$ of the brightest jump operators for $60\times60$ atoms with $d=0.3\lambda_0$ and (bottom) the square root of the determinant of its covariance matrix as a function of the array size $N_\text{1D}$. (c) Top: Plot of relative variance $\Delta$ of the principal eigenvector as a function of the lattice spacing for $N_\text{1D}$ from 20 to 200 in step of 20 (brightest to darkest markers). Bottom: Scaling of $\Delta$ as a function of $N_\text{1D}$. Different markers correspond to different values of $d/\lambda_0$ as specified by the legend.}}
    \protect\label{fig:Delocalization}
\end{figure}

\new{For the scaling law $R_\star \sim N\Gamma_\text{max}$ to hold for atomic arrays, it is necessary for the brightest jump operator $\hat{c}_1$ to be delocalized such that $\Delta\sim O(1)$. As demonstrated in Ref.~\citesupp{Clemens2003collectivesupp}, the collective jump operators of ordered atomic arrays are approximated by Fourier modes of a definite quasi momentum $\bold{k}$ in the limit of $N\gg 1$, and are therefore delocalized. We confirm this in Fig.~\protect\ref{fig:Delocalization}(a,b) (top panels), where we show that the Fourier transform of the brightest collective jump operator is localized (in momentum space) for both 1D and 2D arrays.}
\new{Specifically, we obtain numerically the principal eigenvector $\vec{\alpha}^{(1)}=(\alpha^{(1)}_1,\ldots,\alpha^{(1)}_N)^T$ of $\boldsymbol{\Gamma}$ and compute its momentum space distribution $|\alpha_{\kk}^{(1)}|^2$ where $\alpha_{\kk}^{(1)} = \sum_j^N \exp(\ii \kk \cdot \rr_j) \alpha^{(1)}_j/\sqrt{N}$. Furthermore, we show that the width of $|\alpha_{\kk}^{(1)}|^2$, quantified through the variance $\Delta k^2$, decreases with the size of the array [see bottom panels of Fig.~\protect\ref{fig:Delocalization}(a,b)]. These results are obtained for an array polarized along one of the axis of the array. For perpendicular polarization, a similar narrowing of the distribution in $k$-space is observed albeit with some oscillation modulating the decreasing uncertainty $\Delta k$.}

\new{Finally, to confirm that the upper and lower bounds to $R_\star$ are asymptotically tight, we compute $\Delta$ for a 2D array in the top panel of Fig.~\protect\ref{fig:Delocalization}(c). The values of $\Delta$ are seen to saturate to a finite constant value as $N$ increases. This is confirmed in the bottom panel of Fig.~\protect\ref{fig:Delocalization}(c), where, for a few values of $d$ in the relevant regime $0.1\leq d/\lambda_0 \leq 0.5$, $\Delta$ saturates to a constant value for increasing system sizes.}

\new{For 3D arrays we expect a similar behavior. However,  due to computational constraints on $N_\text{1D}$, we are unable to provide numerical evidence to support this expectation.}\\

\subsection{Effect of position disorder on the scaling laws for atomic arrays}\label{app:Disorder}

Here, we analyze the robustness of the scaling law for arrays, i.e. Eq.~(\ref{eq:array_scaling_law}), to displacement in the atomic positions. Position disorder occurs as either imperfections in the position of the trap (e.g., in tweezer arrays) or temperature fluctuations of the atomic center-of-mass position (e.g., in optical lattices). In both cases, $R_\star$ is obtained from Eq.~(\ref{eq:haux}) where $\boldsymbol{\Gamma}$ is replaced by $\boldsymbol{\Gamma}'$ computed by sampling Eq.~(\ref{eq:Gamma}) according to atoms' position probability distribution. In the following, we consider a Gaussian distribution centered around the lattice sites with a standard deviation $\sigma = \eta d$. If $\eta$ is kept constant, $\sigma$ corresponds to different values of the temperature, or of the tweezers' displacement depending on $d$. Different experiments use different atoms and atomic transitions, resulting in a wide range of values for $\lambda_0$ and hence $d$. In the following, we fix $\eta=0.05$, which corresponds to typical values reported in recent experiments~\citesupp{Rui2020supp,endres2024supp,hofer2024supp}.
\begin{figure}
    \centering
    \includegraphics[width=\columnwidth]{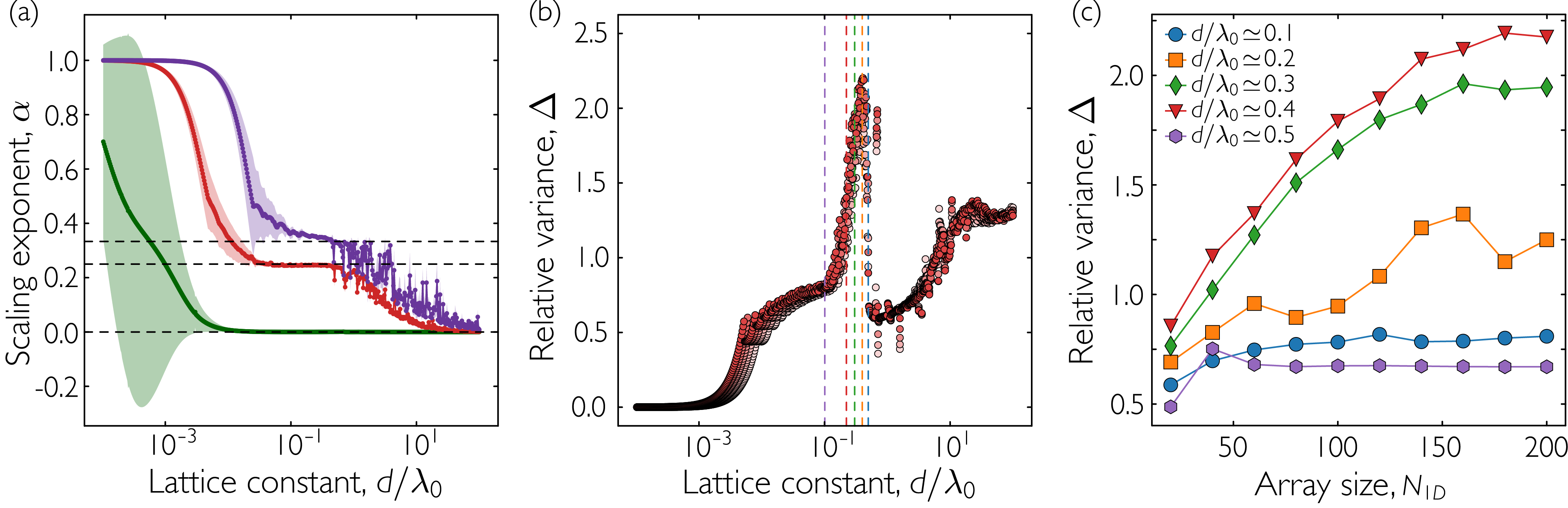}
    \caption{\new{Effects of position disorder on $R_\star$ for arrays of atoms normally distributed around the sites of a square lattice with standard deviation $\sigma=0.05 d$. Atoms are polarized along one of the array axis. All results are obtained by averaging over $100$ position realizations. (a) Scaling exponent $\alpha$ of $\Gamma_\text{max}\sim N^\alpha$ as a function of the average lattice constant $d$. (b) Relative variance $\Delta$ of the principal eigenvector of $\boldsymbol{\Gamma}$ for a 2D array as a function of the lattice spacing $d/\lambda_0$ and for five different values of $N_\text{1D}$ starting from $N_\text{1D}=100$ (lighter color) and increasing in steps of 20 up to $N_\text{1D}=180$ (darker color). (c) Scaling of $\Delta$ with $N_\text{1D}$ for the values of $d/\lambda_0$ corresponding to the vertical dashed line of the same color in panel (b). The exact value of the lattice spacing for each line is specified in the legend.}}
    \protect\label{fig:Disorder}
\end{figure}
We study the effects of disorder on the upper and lower bound to $R_\star$ in Eq.~(\ref{eq:master_bound}) separately. 

The upper bound is entirely determined by the scaling of $\Gamma_\text{max}$. In the limit of weak position disorder ($\eta\ll 1$), the largest eigenvalues of $\boldsymbol{\Gamma}'$ can be analytically computed using perturbation theory. After averaging over the atom position distribution, we obtain (see Appendix C of Ref.~\citesupp{Rusconi2021supp})
\begin{equation}
    \bar{\Gamma}'_\text{max} \simeq (1-\eta^2)\Gamma_\text{max} + \eta^2\Gamma_0.
\end{equation}
We confirm this result by numerically computing the scaling of $\bar{\Gamma}'_\text{max}$ for 1D, 2D, and 3D arrays in Fig.~\protect\ref{fig:Disorder}(a). For 1D arrays, the robustness of the upper bound is sufficient to conclude the validity of the scaling law in Eq.~(\ref{eq:array_scaling_law}) because a lower bound to the decay rate is always given by $N\Gamma_0$ which is achieved by $\ket{e}^{\otimes N}$.

For 2D and 3D arrays, the lower bound depends on both the scaling of $\Gamma_\text{max}$ and of the principal eigenvector of $\boldsymbol{\Gamma}$ through $\Delta$ (see Eq.~\ref{eq:master_bound}). We show numerically that $\Delta\sim O(1)$ for the case of 2D arrays. This is done for 2D arrays in Fig.~\protect\ref{fig:Disorder}(b), where we plot $\Delta$ as a function of $d$ for different array sizes. For fixed $d$, $\Delta$ converges to a constant value as $N$ is increased [Fig.~\protect\ref{fig:Disorder}.(c)]. The numerical results in Figs.~\protect\ref{fig:Disorder}.(b,c) together with the robustness of the scaling of $\Gamma_\text{max}$ allows us to conclude that lower and upper bounds are asymptotically tight leading to the scaling law in Eq.~(\ref{eq:array_scaling_law}) for 2D arrays. For 3D arrays we expect a similar behavior. However,  due to computational constraints on $N_\text{1D}$, we are unable to provide numerical evidence to support this expectation.

\subsection{Peak time of the superradiant burst and Markovianity constraints on system size}\label{app:Burst_Time}

In this appendix, we estimate the timescale $\tau_0$ of the superradiant burst in an atomic array. In analogy with the case of Dicke superradiance~\citesupp{Gross1982supp}, we define $\tau_0$ as the time required to decay from the initial state $\ket{e}^{\otimes N}$, to the state where $R_\star$ is achieved. Since this state is unknown, we estimate $\tau_0$ using the results of Appendix~\ref{appendix:lb}. We approximate the (fastest) decay process as that obtained by successive applications of the jump operator associated to the largest transition rate $\Gamma_\text{max}$. Within this assumption, the time scale is then
\begin{equation}\label{eq:t_burst}
    \tau_0 \equiv \sum_{m=0}^{m_*} \frac{1}{\gamma_m},
\end{equation}
where $\gamma_m$ is given in Eq.~(\ref{eq:gamma_m_Phys_ansatz}) and represents the transition rate from the manifold with $N-m$ excitations to that with $N-m-1$ excitations. The sum runs until $m_*$, which is the optimal number of jumps that maximizes $\gamma_m$, see Eq.~(\ref{eq:m_optimal}). Equation~(\ref{eq:t_burst}) represents an estimation of the shortest time scale for the emission of a burst, as decay through other decay channels with $\Gamma_\mu<\Gamma_\text{max}$ could slow down the radiative emission. To evaluate Eq.~(\ref{eq:t_burst}) we analyze different scenarios depending on the array dimensionality. 

For 2D and 3D arrays in the asymptotic limit $N\rightarrow\infty$, $\Gamma_\text{max}=\beta N^{\alpha}\Gamma_0$ with $\alpha>0$, $m_*=N/2$, and we approximate Eq.~(\ref{eq:gamma_m_Phys_ansatz}) as
\begin{equation}
    \gamma_m \simeq \Gamma_0\frac{(N-m)^2}{N}\pare{1+\frac{m\beta N^\alpha}{N-m}} \simeq N\Gamma_0 \pare{1+m\beta N^{\alpha-1}}.
\end{equation}
For large array size, we approximate the sum in Eq.~(\ref{eq:t_burst}) as an integral and obtain
\begin{equation}
\begin{split}
    \tau_0 &\simeq \frac{1}{N\Gamma_0} \int_{0}^{N/2} \!\!\!\text{d}x\, \frac{1}{1+x\beta N^{\alpha-1}}\simeq \frac{ \log(\beta N^\alpha/2)}{\beta N^\alpha\Gamma_0}.
\end{split}
\end{equation}
In the asymptotic limit $N\rightarrow\infty$, this results holds for any finite value of the interatomic distance $d$.

For 1D arrays the time in Eq.~(\ref{eq:t_burst}) depends on $d$ through $\beta$ even in the asymptotic limit $N\rightarrow \infty$. Indeed, one recovers the Dicke $\tau_0\sim \log(N)/N\Gamma_0$ and non-interacting limits $\tau_0=1/N\Gamma_0$, for $d/\lambda_0\rightarrow 0$ and $d/\lambda_0\rightarrow\infty$, respectively. For independent atoms, the photon emission rate decreases monotonically. Hence, $\tau_0$ represents the average time for the emission of the first photon. In both cases $\beta\rightarrow1$, as shown in Fig.~\protect\ref{fig:Numerical_Scaling}.  In the intermediate region $\beta\gg1$ and $m_*\simeq N/2$, we evaluate Eq.~(\ref{eq:t_burst}) as
\begin{equation}
    \tau_0 \simeq \frac{1}{N\Gamma_0} \int_0^{N/2}\!\!\! \text{d}x\, \frac{1}{1+\beta x/N} = \frac{\log(\beta/2)}{\beta \Gamma_0}.
\end{equation}

Therefore, for the regime in Fig.~\protect\ref{fig:Numerical_Scaling}, for which the scalings of $\beta$ and $\alpha$ are well approximated by the asymptotic values, the timescale of the burst for a $D$-dimensional lattice reads
\begin{equation}\label{eq:t_burst_D}
    \tau_0^{(D)} \simeq \frac{\log[(\beta/2) N^{(D-1)/2D}]}{\beta \Gamma_0 N^{(D-1)/2D}}.
\end{equation}
Comparing this result with that of Dicke, $\tau_0\sim \log(N)/N\Gamma_0$, one finds a generic scaling for the superradiant burst that reads
\begin{equation}
    \tau_0^{(D)} \simeq T_R^{(D)}\log\pare{\frac{1}{2\Gamma_0T_R^{(D)}}},
\end{equation}
where 
\begin{equation}\label{eq:T_R_array}
    T_R^{(D)}\equiv \frac{1}{\beta N^{(D-1)/2D}\,\Gamma_0}
\end{equation}
is the inverse maximal decay rate \textit{per atom}. Equation~(\ref{eq:T_R_array}) agrees with the expression obtained by Arecchi and Courtens~\citesupp{Arecchisupp} for the Dicke limit, which is justified by many authors using different methods (see Ref.~\citesupp{Gross1982supp,Andreev1980supp,Malz2022supp} and references therein). 

Equation~(\ref{eq:T_R_array}) allows us to derive a constraint on the maximum array size $L=N_\text{1D}d$ for which the scaling law in Eq.~(\ref{eq:array_scaling_law}) applies. Our predictions on $R_\star$ rest on the assumption that the Born-Markov master equation~(\ref{eq:lindbladME}) correctly describes the dynamics of the system. It is thus necessary for the inverse maximal decay per atom, $T_R^{(D)}$ in Eq.~(\ref{eq:T_R_array}), to satisfy the conditions $T^{(D)}_R\gg \tau_c$ and $T^{(D)}_R\gg L/c$, where $\tau_c\sim 1/\w_0$ is the correlation time of the electromagnetic field bath and $L/c$ is the propagation time of a photon across the system~\citesupp{Lehmberg1970asupp,Lehmberg1970bsupp}. Of the two conditions, the latter one (that is, to ignore retardation) is more stringent for extended atomic ensembles in free space and imposes a constraint on the (lateral) system size $L=N_\text{1D}d$. Hence, we estimate the limits on the number of atoms as 
\begin{equation}\label{eq:Limit_N1D_general}
    N_\text{1D} \ll \bigg(\frac{\w_0}{\beta \Gamma_0} \frac{1}{k_0 d}\bigg)^{\frac{1}{\alpha D +1}},
\end{equation}
where, for a finite atomic array, $\alpha$ and $\beta$ depend on the lattice constant as discussed in Appendix~\ref{app:scaling_Gamma_max}. For $0.01 \lesssim k_0 d\lesssim 1$, the behaviour of $R_\star$ in finite arrays is well captured by the asymptotic scaling for $N\gg 1$, and approaches the non-interacting limit for $k_0 d\gtrsim 1$. Accordingly, we can write Eq.~(\ref{eq:Limit_N1D_general}) as
\begin{equation}\label{eq:Limit_N1D_Asymptotic}
    N_\text{1D} \ll \Big(\frac{\w_0}{\Gamma_0}\Big)^{\frac{2}{D+1}}f(k_0 d)
\end{equation}
where
\begin{equation}\label{eq:Limit_N1D}
    f(x) = \left\{
    \begin{array}{ll}
    x^{\frac{D-1}{D+1}}&\quad \text{for}\, x\lesssim 1,\\
    x^{-1}&\quad \text{for}\, x\gtrsim 1.
    \end{array}
    \right.
\end{equation}
Equation~(\ref{eq:Limit_N1D_Asymptotic}) recovers the expected inverse scaling with large lattice constants. At subwavelength interatomic spacing, the maximal decay rate $R_\star$ decreases with increasing $k_0d$, thus allowing for larger number of atoms before breaking Markovianity. This is particularly relevant for 1D arrays, where the dependence of $T^{(D)}_R$ and $L/c$ on $d$ cancels exactly. For atomic transitions in the optical domain, $\w_0/\Gamma_0 \simeq 10^8$, Eq.~(\ref{eq:Limit_N1D}) yields the asymptotic conditions for $N_\text{1D} \ll 10^8$,  $N_\text{1D} \ll 10^5 (k_0 d)^{1/3}$, and $N_\text{1D} \ll 10^4 (k_0d)^{1/2}$ for one-, two-, and three-dimensional subwavelength arrays, respectively.

% ======================================
\section{Experimental implications of the scaling laws}

In this section, we expand the discussion on the experimental implications of the bound $R_\star$ presented in the main text. In particular, we discuss transient superradiance in Sec.~\ref{app:g2}, driven-dissipative phase transitions in atomic arrays (Sec.~\ref{sec:SR_Laser} and \ref{sec:Coherent_Pumping}), and error rates of quantum simulators and processors based on Rydberg atom arrays (Sec.~\ref{app:Rydberg_Collective}).

\subsection{Transient superradiance: relation between $g^{(2)}(0)$ and $R_\star$}\label{app:g2}

The scaling laws of $R_\star$ in Eq.~\eqref{eq:array_scaling_law} determine a rigorous upper bound on the maximum scaling achievable by superradiant decay in free space arrays. There is however no guarantee that $R_\star$ is achieved during dynamical evolution. A separate condition relates the early time correlation to the appearance of a superradiant burst, i.e. a non monotonic evolution of $R(t)$. We reveal a connection between the correlations at early time and the scaling of $R_\star$.

It was recently demonstrated that atomic arrays prepared in the state $\ket{e}^{\otimes N}$ display a burst if the emission of the first photon in a decay channel enhances the probability of a second photon being emitted in the same channel, i.e., if $g^{(2)}(0)>1$~\citesupp{Masson2020supp,Masson2022NatCommsupp,Robicheaux2021supp,mok2023dickessupp}. We can rewrite this condition in terms of
\begin{equation}\label{eq:g2_condition}
    \dot{R}(t=0) = \Vert \mathbf{\Gamma}\Vert_F^{2} - 2N\Gamma_0^2,
\end{equation}
where $\Vert \mathbf{\Gamma}\Vert_F = \text{Tr}(\mathbf{\Gamma}^2)$ is the Frobenius norm of the decoherence matrix $\mathbf{\Gamma}$. The system will display a superradiant burst (at short times) if $\dot{R}(t=0) > 0$. The converse is not proven, but can be argued using second-order mean-field theory, see Theorem 2 in the Supplementary Information of Ref.~\citesupp{mok2023dickessupp}. Making use of norm inequalities and the bounds on $R_\star$ presented in the main text, we can bound $\dot{R}(t=0)$ in terms of $R_\star$. For example, from Eq.~(\ref{eq:master_bound}), and noting that $\Vert \mathbf{\Gamma} \Vert_F \leq \sqrt{\text{rank}(\mathbf{\Gamma}}) \Gamma_{\text{max}}$~\citesupp{Bhatia1997matrixsupp}, we find
\begin{equation}\label{eq:Bound_g2_Rmax}
    \dot{R}(t=0) \leq \frac{16}{N^2} (1+\Delta^2)^2 \text{rank}(\mathbf{\Gamma}) R_\star^2 - 2 N\Gamma_0^2.
\end{equation}
This is asymptotically tight and saturated in the Dicke limit. 

One corollary is a no-go result on the existence of a superradiant burst: if $8(1+\Delta^2)^2 R_\star^2 < N^3\Gamma_0^2/\text{rank}(\mathbf{\Gamma})$, then a burst cannot occur. In contrast, using the general upper bound $R_\star \leq N(3\Gamma_{\text{max}}-\Gamma_0)/2$ and noting that $\Gamma_{\text{max}} \leq \Vert \mathbf{\Gamma} \Vert_F$, we derive
\begin{equation}
    \dot{R}(t=0) \geq \frac{1}{9}\left(\frac{2R_\star}{N\Gamma_0} + 1\right)^2 - 2N\Gamma_0,
\end{equation}
which guarantees a superradiant burst if $R_\star/\Gamma_0 \gtrsim 3N^{3/2}/\sqrt{2}$.

\subsection{Incoherently driven arrays: a free space superradiant laser?}\label{sec:SR_Laser}

Here we analyze the problem of incoherently pumped atoms, where the dynamics is described by Eq.~(\ref{eq:lindblad_ME_incoher_drive}), setting $\eta_{j}=0$. Atoms are assumed to be independently pumped, i.e., $W_{ij}=W\delta_{ij}$. For atoms in a bad cavity -- when the cavity linewidth exceeds the atomic linewidth -- the system undergoes a phase transition into a superradiant lasing regime where atoms spontaneously emit coherent light. This occurs for pump strengths $\Gamma_0\lesssim W\lesssim R_\star/N$, where $R_\star = N^2\Gamma_0/4$, and the emitted light intensity scales as $I_\text{max}\sim R_\star/N$ when maximized over $W$~\citesupp{Meiser2009supp,bohnetsupp}. It is unknown whether superradiant lasing can occur beyond the cavity paradigm discussed in the literature. 

Below, we demonstrate that the maximum light intensity (as well as the optimal pump strength $W_\star$) in the superradiant lasing region -- \emph{if} such regime exists for ordered arrays in free space -- are both upper-bounded by $R_\star$. The rate of change of the number of excitations is
\begin{equation}
    R =\sum_{i,j=1}^N\Gamma_{ij}\avg{\sp_j\sm_i} - W \sum_{j=1}^N \avg{\sm_j\sp_j}.
\end{equation}
The first term on the right hand side corresponds to the rate of free-space emission and contributes to the light intensity [see Eq.~(\ref{eq:Intensity_direction})]. The second term, instead, corresponds to the rate of change due to the external pump. In the steady state $R=0$, and thus
\begin{equation}
    \sum_{j=1}^N\avg{\sp_j\sm_j}W = \sum_{i,j=1} ^N\Gamma_{ij}\avg{\sp_j\sm_i}\leq R_\star.
\end{equation}
For 2D and 3D arrays, as well as for 1D array in the region $10^{-3}\lesssim d/\lambda_0\lesssim 10^{-1}$ (see discussion in Sec.~\ref{appendix:lb}), the emission rate is maximized in the manifold containing $N/2$ excitations. We thus obtain 
\begin{equation}\label{eq:W_star}
    W_\star \leq \frac{2R_\star}{N}. 
\end{equation}
This bound is tight for atoms in a cavity (Dicke limit). Combining the scaling of the optimal pump strength in Eq.~(\ref{eq:W_star}) and that of the maximal emitted intensity, the scaling of the superradiant lasing region for atomic arrays in free space follows the scaling laws in Eq.~(\ref{eq:array_scaling_law}). This result indicates the impossibility of observing a superradiant phase transition in a 1D atomic array in free space, as the intensity never scales superlinearly with the system size.

\subsection{Coherently driven atomic arrays: Collective resonance fluorescence}\label{sec:Coherent_Pumping}

We now consider the problem of a coherently driven atomic array. The master equation describing the evolution is obtained from Eq.~(\ref{eq:lindblad_ME_incoher_drive}) by setting $W_{ij}=0$. This master equation describes collective resonance fluorescence from an ensemble of dipole-interacting atoms in free space~\citesupp{Ostermann2024supp,Agarwal2024supp,Goncalves2024supp,Ruostekoski2024supp}.

In the Dicke limit (i.e., for atoms in a cavity), the system undergoes a second order phase transition as the Rabi frequency of the drive $\eta$ is increased above a critical threshold $\eta_c=N\Gamma_0/2$~\citesupp{narduccisupp,carmichael_1980supp}. Below threshold, $\eta<\eta_c$, atoms are in a magnetized phase, with a non-zero average dipole moment whose magnitude is determined by the balance between the drive strength and spontaneous emission. At threshold, the drive strength equals the maximum emission rate (per atom). Further increasing the drive strength saturates the atoms, destroying coherence between them, leading to a paramagnetic phase with zero average dipole moment.

We demonstrate below that, for arrays of atoms in free space, both the scaling of critical drive strength and of the maximum radiated intensity in the paramagnetic phase are bounded by $R_\star$. 
% We define the critical pump strength as the rate at which the Rabi frequency balances the decay rate of the atomic coherence $\hat{S}_y = \sum_j\hat{\sigma}^y_j/2$.
% Using the bound for the rate of change of general observables in Eq.~(\ref{eq:localop_Rstarbound}), we obtain
% \begin{equation}
%     \bigg|\frac{\text{d}}{\text{d}t}\langle \hat{S}_y\rangle\bigg| \leq N\sqrt{\Gamma_0 R_\star},
% \label{eq:Bound_Decay_Coherence}
% \end{equation}
% where the factor of two in Eq.~(\ref{eq:localop_Rstarbound}) has been canceled by the factor $1/2$ appearing in the definition of $\hat{S}_y$. Assuming a spatially-uniform drive, the rate of change of $\avg{\hat{S}_y}$ caused by the driving Hamiltonian $\hat{V}_\eta$ is upper bounded by $\eta N/2$. From Eq.~(\ref{eq:Bound_Decay_Coherence}), we estimate an upper bound on the critical pump strength 
% \begin{equation}\label{eq:Critical_eta_general}
%     \eta_c \leq 2\sqrt{\Gamma_0 R_\star}.
% \end{equation}
% This result does not make any assumption on the state of the system achieving the bound, and it is thus valid in general. It does assume, however, that the rate of change due to the coherent evolution has a contribution smaller than $R_\star$. Equation~(\ref{eq:Critical_eta_general}) correctly captures the scaling of $\eta_c$ for the Dicke model but it overestimates its exact expression by a factor of two.
We assume a mean field ansatz for the state of the system, as done in recent work~\citesupp{Ostermann2024supp,Goncalves2024supp,Agarwal2024supp,Ruostekoski2024supp}. Specifically, we assume the product state $\hat{\rho}_\text{ss} = \bigotimes_j \hat{\rho}_\text{ss}^j$, where $\hat{\rho}_\text{ss}^j$ is the steady state of a single atom. This is expected to be a good approximation at large pump strengths, as the coherences between the atoms are lost as they saturate. The critical pump strength $\eta_c$ can thus be obtained from the steady state condition $R=0$, where 
\begin{equation}\label{eq:Rate_Coher_Driving}
    R = \sum_{i,j=1}^N\Gamma_{ij}\avg{\sp_j\sm_i} - \ii \sum_{j=1}^N \Big(\eta_j \avg{\sm_j} - \eta^*_j \avg{\sp_j} \Big)
\end{equation}
is the instantaneous rate of change of the average number of excitation derived from Eq.~(\ref{eq:lindblad_ME_incoher_drive}). Evaluating Eq.~(\ref{eq:Rate_Coher_Driving}) on the product state $\hat{\rho}_\text{ss}$, and assuming a resonant drive with spatial dependence $\eta_j = \eta e^{i \kk_0\cdot\rr_j}$, we obtain, in the limit $\eta\gg\Gamma_0$,
\begin{equation}\label{eq:eta_c}
    \eta_c = \frac{\Gamma_0}{2}\sqrt{1+\frac{\Gamma_{\text{max}}}{\Gamma_0}} \approx \frac{\Gamma_0}{2}\sqrt{\frac{R_\star}{N\Gamma_0}}.
    % \sim \Gamma_0 \sqrt{\frac{\Gamma_\text{max}}{\Gamma_0}}.
\end{equation}
This result calls for several remarks.
% First, Equation~(\ref{eq:eta_c}) does not yield the correct values of $\eta_c$ for the Dicke limit. This is not surprising as the ansatz state we used to arrive to Eq.~(\ref{eq:eta_c}) is not permutation invariant.
First, the suppression of the single atom coherences in the high-saturation regime does not necessarily imply the absence of correlations between atoms. Hence, a mean field treatment is not a priori justified. In this context, however, this approximation has been validated in numerous recent works~\citesupp{Ostermann2024supp,Goncalves2024supp,Agarwal2024supp,Ruostekoski2024supp}. Second, as shown by our proof, the key ingredient for saturating $R_\star$ (up to a constant factor) is the phase coherence of the quantum state relative to the phases imprinted on the atoms by the dominant collective jump operator $\hat{c}_1$. Accordingly, the assumption to drive the system at the light cone (that is, with $\eta_j = \eta e^{\ii \kk_0\cdot\rr_j}$), is crucial to obtain the dependence on $R_\star$. We stress that this occurs naturally when the drive is directed along the array direction in 1D, in the array plane in 2D, and along any direction in 3D. This connects our results with previous work on collective resonance fluorescence in free space arrays. In particular, the phase matching condition explains the absence of any system-size scaling for 2D arrays driven perpendicularly to the array plane~\citesupp{Ostermann2024supp,mi2024stablesupp}. Third, the square-root dependence on the rate $R_\star$ is analogous to the dependence $\eta_c\sim\Gamma_0\sqrt{N}$ obtained in Ref.~\citesupp{Agarwal2024supp} for collective resonance fluorescence in an elongated (disordered) cloud if one assumes $R_\star = N^2\Gamma_0$ in Eq.~(\ref{eq:eta_c}). This suggests that the scaling $\sim\sqrt{R_\star/N}$ is a general feature arising from the extended nature of the system, irrespective of the arrangement of the atoms.
% Fourth, the argument leading to Eq.~\eqref{eq:eta_c} does not make any assumptions on the nature of the Hamiltonian interactions in Eq.~\eqref{eq:lindbladME}, in contrast with the bound derived in Eq.~\eqref{eq:Critical_eta_general}.
Finally, it is unclear whether the crossover at $\eta=\eta_c$ is a true phase transition, with work on atomic clouds suggesting that it is not~\citesupp{Goncalves2024supp,Agarwal2024supp}. Nevertheless, the scattered light in the two regimes has distinctive scalings with the system size, as we now discuss.

The light intensity emitted by the system of $N$ atoms along a direction $\nn$ reads~\citesupp{dicke1954coherencesupp}
\begin{equation}\label{eq:Intensity_direction}
    I(\nn,t) = I_0 D(\nn) \sum_{i,j=1}^N e^{\ii k_0\nn\cdot (\rr_i-\rr_j)}\avg{\sp_j\sm_i},
\end{equation}
where $I_0=\hbar \w_0 \Gamma_0$ is the intensity emitted by a single atom and we define the dipole emission pattern $D(\nn)$ such that $\int\!\text{d}\nn\,D(\nn)=1$. Assuming the system is in the steady state $\hat{\rho}_\text{ss}$, the total integrated intensity reads
\begin{equation}\label{eq:Integrated_Intensity}
    \frac{I}{I_0} = N\avg{\sp\sm}_\text{ss} + N|\avg{\sp}_\text{ss}|^2 \mu,
\end{equation}
where $\avg{\sp\sm}_\text{ss}=2\eta^2/(\Gamma_0^2+4\eta^2)$ and  $|\avg{\sp}_\text{ss}|^2=4\eta^2 \Gamma_0^2/(\Gamma_0^2+4\eta^2)^2$ are the steady state population and coherence of a resonantly driven single atom. We also introduced the dimensionless factor
\begin{equation}\label{eq:function_f}
    \mu \equiv -1 + \frac{1}{N}\int\!\!\text{d}\nn\, D(\nn)\sum_{i,j=1}^N e^{\ii (k_0\nn-\kk_0)\cdot (\rr_j-\rr_i)},
\end{equation}
which depends on the array shape and size and describes the effects of interference on the emitted light intensity. Evaluating the discrete sum in Eq.~(\ref{eq:function_f}), in the limit of large arrays ($N\gg1$) one can prove that $N\mu \propto N^{\frac{3}{2}-\frac{1}{2\text{D}}}$, for arrays with constant density [see for instance Eq.~(C6) in Ref~\citesupp{Abella1966supp}].

From this result it is easy to take the limit of weak and strong pump strength with respect to $\eta_c$. For weak driving, we obtain 
\begin{equation}\label{eq:Intensity_weak_Driving}
    \frac{I}{I_0} \approx \frac{2\eta^2}{\Gamma_0^2}N + \frac{4\eta^2}{\Gamma_0^2}N\mu \sim O\pare{\frac{R_\star}{\Gamma_0}} \qquad (\eta\ll\Gamma_0),
\end{equation}
This result confirms the superlinear scaling of the intensity in the weak driving regime predicted in Ref.~\citesupp{Ruostekoski2024supp}. Furthermore, it predicts a precise value for maximum achievable scaling exponent in the mean field approximation. For pump strengths above the saturation threshold, we obtain
\begin{equation}\label{eq:Intensity_strong_Driving}
    \frac{I}{I_0} \approx \frac{N}{2} + \frac{\Gamma^2_0}{4\eta^2}N\mu \sim O(N),\qquad (\eta>\eta_c),
\end{equation}
where we used Eq.~(\ref{eq:eta_c}) and assumed that $\eta\propto \eta_c$ with a proportionality factor that does not scale with $N$. Our results are consistent with  Ref.~\citesupp{Agarwal2024supp}, where the scaling of the directional emission by an elongated cloud of atoms is computed within the mean field approximation. The directional emission from an elongated cloud exhibits superlinear $N$-scaling at small pump rate ($\eta\ll\eta_c$), which changes to a linear scaling $N$ above threshold.

\subsection{Quantum computing and simulation: Rydberg collective decay}\label{app:Rydberg_Collective}

In this section, we consider the effect of collective decay in Rydberg arrays. The problem of Rydberg collective decay has been amply analyzed~\citesupp{Suarez2022supp,Grimes2017supp} but, outside of the cavity scenario, theoretical analysis have been restricted to Dicke-limit approximations for dense atomic clouds~\citesupp{Wang2007supp,Hao2021supp}. We analyze the collective decay rate via the scaling laws derived before, drawing conclusions that hold for a large range of system sizes. 

We focus on setups based on $^{87}\text{Rb}$ atoms in tweezer arrays from Refs.~\citesupp{Ebadi2021supp,Manovitz2025supp,Evered2023supp,Bluvstein2024supp}, where a large fraction of atoms in the array is excited to a Rydberg level to either generate relevant many-body states or to perform parallel gates. Qubits are encoded in the hyperfine ground-state levels $|0\rangle=|5S_{1/2},F=1,m_F=0\rangle$ and $|1\rangle=|5S_{1/2},F=2,m_F=0\rangle$, driven to the Rydberg level $|r\rangle=|53S_{1/2},m_J=+1/2\rangle$ via a two-photon transition through the intermediate level $|l\rangle=|6P_{3/2},F=1\rangle$ (as in Ref.~\citesupp{Evered2023supp}, we take single-photon Rabi frequencies $\Omega_{420}=2\pi\times 237\,\text{MHz}$, $\Omega_{1013}=2\pi\times 303\,\text{MHz}$, intermediate state detuning $\Delta=2\pi\times 7.8\,\text{GHz}$, and two-photon Rabi frequency $\Omega=\Omega_{420}\Omega_{1013}/2\Delta=2\pi\times 4.6\,\text{MHz}$). Atoms in the Rydberg state interact via both short-range Casimir ($V_{nn}=C_6/d^6$, with $C_6=2\pi\times 28.8\,\text{GHz}\,\mu\text{m}^6$) and long-range dipole-dipole interactions. The first is of coherent nature and gives rise to an effective blockade radius $R_b=(C_6/\Omega)^{1/6}=4.29\,\mu\text{m}$ that shifts the Rydberg levels of nearby atoms preventing them from getting excited. The former has both a coherent and dissipative contribution that might manifest in the form of collective frequency shifts and collective dissipation. 

\begin{figure}
    \centering
    \includegraphics[width=\linewidth]{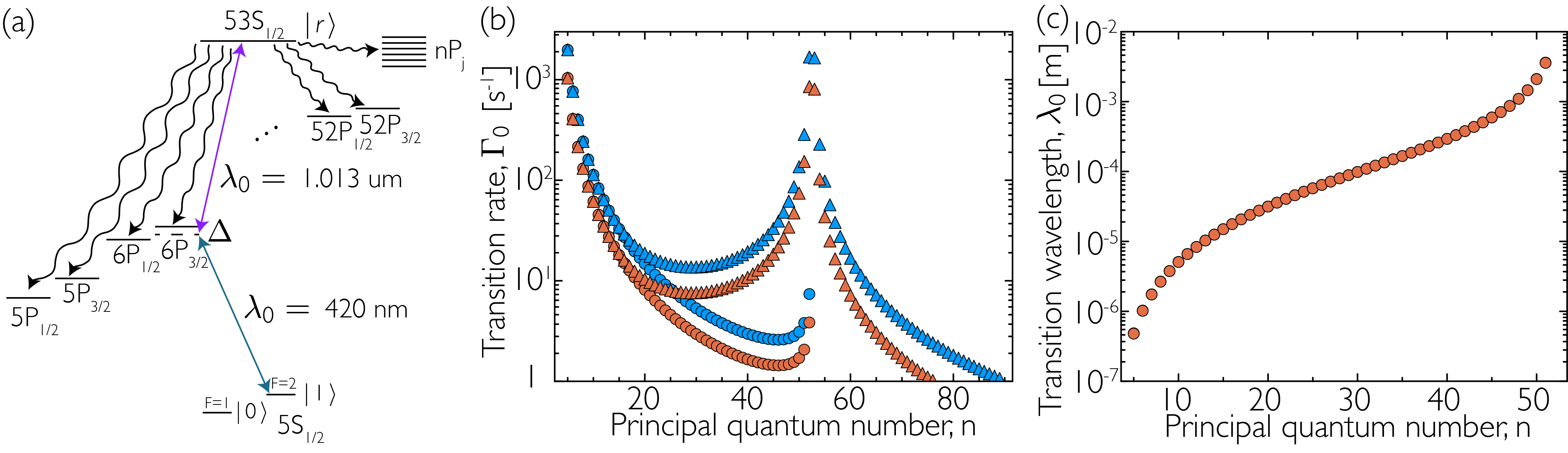}
    \caption{\new{Decay from the Rydberg level $53S_{1/2}$ of $^{87}\text{Rb}$. All possible decay paths are shown in (a), including black-body enhanced transitions to higher-energy states ($nP_j$, with $n\geq 53$), in addition to the qubit storage ground-state levels and the two-photon transition used. (b) Atoms in the Rydberg level can decay both to levels with $J=1/2$ (shown in orange) and $J=3/2$ (shown in blue). Transition rates are shown for $T=0\text{ K}$ (circles) and $T=300\text{ K}$ (triangles), accounting for black-body enhancement. (c) Transition wavelengths. The $J=3/2$ data is not visible due to an almost perfect overlap with $J=1/2$. All data has been calculated using the ARC package~\protect\citesupp{ARCsupp,Beterov2009supp}.}}
    \protect\label{fig:Rydberg}
\end{figure}

We consider all possible direct decay paths from $|r\rangle$ (i.e., all transitions to levels $iP_{1/2}$ and $iP_{3/2}$, with $i\geq 5$), as summarized in Fig.~\protect\ref{fig:Rydberg}(a). While only transitions to $i\leq 52$ levels exhibit spontaneous decay, these transitions as well as transitions to levels with higher principal quantum number can be black-body enhanced [Fig.~\protect\ref{fig:Rydberg}(b)]. We focus our attention on $iP_{3/2}$ levels, as they dominate over $iP_{1/2}$ due to the slightly longer wavelengths and (twice) larger decay rates [Figs.~\protect\ref{fig:Rydberg}(b)-(c)]. The multilevel nature of the system can be simplified for large atom numbers, as the dominant decay channel will overcome all other possible decay paths~\citesupp{Masson2024supp, Mok2025supp}. We find that, for array sizes such that $L=N_\text{1D}d\leq\lambda_{53S_{1/2}\to 52P_{3/2}}$, the main contribution to collective decay arises only from Dicke-like decay on the $|r\rangle\to|52P_{3/2}\rangle$ transition.

The relevant scaling for $R_\star$ is determined by the ratio between the lattice constant and the transition wavelength. The $|r\rangle\to|52P_{3/2}\rangle$ transition is in the Dicke regime with maximum decay rate $R_\star = N(N+2)\Gamma_0/4$ when $L\leq \lambda_{53S_{1/2}\to 52P_{3/2}}$, where for simplicity we have defined $\Gamma_0\equiv \Gamma_0^{53S_{1/2}\to 52P_{3/2}}$. Decay predominantly occurs on the $|r\rangle\to|52P_{3/2}\rangle$ transition when its collective decay rate dominates over all other decay channels $|r\rangle\to|iP_{3/2}\rangle$ (which may also be collectively enhanced, depending on the transition wavelength). We classify other decay channels in three groups: Dicke-enhanced transitions (which occurs to other nearby Rydberg levels), other collectively-enhanced transition beyond the Dicke regime where the scaling is given by Eqs.~(\ref{eq:expansion_2D})-(\ref{eq:expansion_3D}), and independent decay channels for which no collectively enhancement occurs. Because the transition rate has a non-monotonic behavior [Fig.~\protect\ref{fig:Rydberg}(b)], and due to the strong transition-wavelength dependence on the principal quantum number [Fig.~\protect\ref{fig:Rydberg}(c)], we conclude that $|r\rangle\to|52P_{3/2}\rangle$ dominates the decay process whenever 
\begin{align}\label{eq:coll}
    &\Gamma^{53S_{1/2}\to52P_{3/2}}_\text{max}\geq \underset{5\leq i \leq 52}{\text{max}} \left(\bar{n}_{53S_{1/2}\to iP_{3/2}}+1\right)\Gamma_0^{53S_{1/2}\to iP_{3/2}}\nonumber\\
    &\iff N\geq\frac{4\left(\bar{n}_{53S_{1/2}\to 5P_{3/2}}+1\right)\Gamma_0^{53S_{1/2}\to 5P_{3/2}}}{\Gamma_0}\approx\frac{4\Gamma_0^{53S_{1/2}\to 5P_{3/2}}}{\Gamma_0},
\end{align}
where we have included black-body enhanced decay with $\bar{n}_{53\to i}=(e^{\hbar\omega_{53\to i}/k_BT}-1)^{-1}$ given by the Planck distribution at temperature $T$, and $\omega_{53\to i}$ is the transition frequency.

Collective decay becomes increasingly relevant when it is faster than any other decay rate. We define the ``individual'' contribution to decay as
\begin{equation}
    \Gamma_\text{ind}\equiv \Gamma_\text{bbr} + \Gamma_\text{sp}=\sum_{\substack{i\geq 5 \\ j\in\left\{1/2,3/2\right\}}}\bar{n}_{53S_{1/2}\to i P_j}\Gamma_0^{53S_{1/2}\to i P_j} + \sum_{\substack{5\leq i \leq 52 \\ j\in\left\{1/2,3/2\right\}}}\Gamma_0^{53S_{1/2}\to i P_j}.
\end{equation}
We simplify the multilevel picture by assuming  that collective decay is only present in the dominant transition and all other transitions decay independently. The maximal decay (per atom) can be approximated as
\begin{equation}
    \frac{R_\star}{N} \simeq \Gamma_\text{tot} \equiv \Gamma^{53S_{1/2}\to52P_{3/2}}_\text{max} + (\Gamma_\text{ind}-\Gamma_0),
\end{equation}
where decay occurs individually over all transitions $i\neq 52$ and collectively for the level $i=52$. Accordingly, we removed the individual decay over the transition $53S_{1/2}\rightarrow 52P_{3/2}$ from $\Gamma_\text{ind}$. Then, the ratio between the maximal decay rate per atom and the nearest-neighbor Casimir interaction rate is
\begin{equation}
    \chi\equiv\frac{R_\star/N}{V_\text{nn}}\simeq\frac{\Gamma_\text{tot}}{V_\text{nn}} =\frac{(N-2)\Gamma_0/4 + \Gamma_\text{ind}}{C_6/d^6}.
\end{equation}
The value of $\chi$ for relevant system sizes is reported in Fig.~\protect\ref{fig:Applications}(b) in the main text. This analysis can be easily extended to other scenarios were the dominant transition lies in the regime of array-like scaling, depending on the choice of atom and the Rydberg level used.

There are two restrictions on the achievable system sizes. The first one is determined by an objective aperture limit of $\text{NA} = 0.65$~\citesupp{endres2024supp}. This limits the objective field of view to $1.5\,\text{mm}$, and therefore restricts tweezer arrays to have a maximum lateral size of $L\leq 1 \text{mm}$ with current technological capabilities [see Fig.~\protect\ref{fig:Applications}(b)]. The second is given by Markovianity, which imposes $N\ll (c/\Gamma_0 d)^{2/3}$, to ensure that the time it takes a photon to travel the entirety of the system is significantly smaller than the smallest possible dissipative timescale of the system. This limit does not show up in Fig.~\protect\ref{fig:Applications}(b) as it only becomes relevant for very large systems.\\

{\bf Quantum computing.} 
One of the recent experimental advances towards neutral-atom quantum computing consists on generating high-fidelity entangling gates in a parallel fashion by creating a tweezer array of groups of atoms (dimers for two-qubit gates)~\citesupp{Evered2023supp}. To apply parallel entangling two-qubit gates, each qubit pair has to be at a distance $d_b<R_b$, with distance between pairs $d>R_b$. We consider the setup described in Ref.~\citesupp{Bluvstein2024supp}, where the entangling process is performed with a $4\times20$ array of pairs (160 atoms in total) at $d=12\,\mu\text{m}$ with $d_b=2\,\mu\text{m}$. At that pair distance, the Casimir interaction rate is $V_\text{nn}=C_6/d_b^6=2\pi\times 450\,\text{MHz}$, orders of magnitude larger than any other rate in the system. We exactly diagonalize the dissipative matrix and find $\Gamma_{\text{max}}^{53\to52}=2\pi\times176\,\text{Hz}$, a factor of two smaller than the largest single atom decay rate $\Gamma_0^{|r\rangle\to 5P_{3/2}}=2\pi\times 315\,\text{Hz}$, and over an order of magnitude smaller than the total single-atom decay $\Gamma_\text{ind}= 2\pi\times3.22\,\text{kHz}$. Using the exact result for $\Gamma_{\text{max}}^{53\to52}$, we find $\chi\approx (\Gamma_{\text{max}}^{53\to52}/4+\Gamma_\text{ind})/V_\text{nn}=7.25\times10^{-6}$, thus concluding that for current state-of-the-art setups, the main error arises from black-body enhanced decay, not collective decay.

The gate error is given by $1-F \simeq \Gamma_\text{tot} T_\text{g}$, where $F$ is the gate fidelity and $T_\text{g}$ the time duration of the gate, which is proportional to the inverse of the two-photon Rabi frequency $\Omega$. We consider the Jandura-Pupillo gate with an optimal gate time $T_\text{g}^\star \simeq 2.95/\Omega$~\citesupp{Jandura2022supp}. Within these assumptions, the gate error is
\begin{equation}
    \varepsilon_{4\times20}=1-F\approx 2.95\,\chi\,\frac{V_\text{nn}}{\Omega}\approx 0.2\%.
\end{equation}

The importance of collective decay increases with system size. To exemplify this, we consider the same scenario but now with a $40\times200$ array of pairs (16000 atoms in total). The exact value for the collective decay rate is now $\Gamma_\text{max}=2\pi\times 6.8\,\text{kHz}$, which leads to a maximum gate error of 
\begin{equation}
    \varepsilon_{40\times200}=1-F\approx 0.3\%,
\end{equation}
meaning that collective decay accounts for $\sim1/3$ of the dissipative contribution to the gate error. This assumes that the same Rabi frequency can be achieved for such a setup and shows that, if all other errors are kept constant, collective decay could become the main source of error, with an error per gate that scales with the number of parallel gates. Nevertheless, these are rough error estimates, and a more detailed analysis should be performed.\\

{\bf Quantum simulation.} We now consider a similar setup to that in Ref.~\citesupp{Manovitz2025supp} to study quantum phase transitions, with a $16\times16$ 2D array of lattice constant $d=R_b/1.1$. During the experiment, half of the atoms are kept in the Rydberg state $|r\rangle$ for a total time of $V_\text{nn}t\approx 110$. The exact collective decay rate of the $|r\rangle\to |52P_{3/2}\rangle$ transition is $\Gamma_{\text{max}}^{53\to52}=2\pi\times301\,\text{Hz}$, an order of magnitude smaller that the total decay rate due to all independent decay paths. This yields $\chi\approx (\Gamma_{\text{max}}^{53\to52}/4+\Gamma_\text{ind})/V_\text{nn}=4.02\times10^{-4}$. This number increases with system size. For a $100\times100$ array, the rate goes up to $\Gamma_{\text{max}}=2\pi\times11.7\,\text{kHz}$, already being the dominant decay rate and therefore opening a possibility of losing a considerable amount of Rydberg atoms during an experimental run. In particular, $\chi\approx7.5\times10^{-4}$, so the probability for each atom to decay from the Rydberg state is  $P\approx 1-e^{-\chi V_\text{nn}t} \simeq 8\%$, doubling the probability obtained from considering individual decay only. All parameters used for this calculation have been rescaled from those in Ref.~\citesupp{Manovitz2025supp}, as that experiment is performed using the $\ket{r}=\ket{70S_{1/2}}$ Rydberg level.

\vspace{.5cm}
\bibliographystylesupp{apsrev4-2}
\bibliographysupp{supp}

\end{document}